# Identification and Quenching of Nugget Galaxies in the RESOLVE Survey at $z=0$

Derrick S. Carr[1], Sheila J. Kannappan[1], Mark A. Norris[2], Manodeep Sinha[3,4], Michael L. Palumbo, III[5],
Kathleen D. Eckert[1], Amanda J. Moffett[6], Mugdha S. Polimera[1], Joel I. Bernstein[1], Zackary L. Hutchens[1], and David V. Stark[7]
[1] Department of Physics & Astronomy, University of North Carolina at Chapel Hill, Chapel Hill, NC 27599, USA
[2] Jeremiah Horrocks Institute, University of Central Lancashire, Preston, PR1 2HE, UK
[3] SA 118, Center for Astrophysics & Supercomputing, Swinburne University of Technology, 1 Alfred Street, Hawthorn, VIC 3122, Australia
[4] ARC Centre of Excellence for All Sky Astrophysics in 3 Dimensions (ASTRO 3D), Australia
[5] Department of Astronomy & Astrophysics, 525 Davey Laboratory, The Pennsylvania State University, University Park, PA 16802, USA
[6] Department of Physics and Astronomy, University of North Georgia, 3820 Mundy Mill Road, Oakwood GA 30566, USA
[7] Space Telescope Science Institute, Baltimore, MD 21218, USA
Received 2023 December 20; revised 2024 March 22; accepted 2024 April 10; published 2024 June 5

## Abstract

We present a complete census of candidate nuggets, i.e., dense galaxies likely formed by compaction with intense gas influx, within the volume-limited redshift $z \sim 0$ REsolved Spectroscopy Of a Local VolumE (RESOLVE) survey. These nuggets span all evolutionary stages and 3 orders of magnitude in stellar mass ($M_* \sim 10^8$ to $10^{11} M_\odot$) from the dwarf to the giant regime. We develop selection criteria for our $z \sim 0$ nugget candidates based on structure and introduce the use of environmental criteria to eliminate nugget-like objects with suspected non-compaction origins. The resulting $z \sim 0$ nuggets follow expectations with respect to structure (i.e., density, size), population frequency, and likely origins. We show that the properties of our nugget census are consistent with permanent quenching above the gas-richness threshold scale (halo mass $M_{\rm halo} \sim 10^{11.4} M_\odot$), cyclic temporary quenching below the threshold scale, and feedback from active galactic nuclei (AGN) assisting in permanent quenching. As predicted in simulations, most nuggets quench within a halo mass range of $M_{\rm halo} \sim 10^{11.45}$ to $10^{11.9} M_\odot$. We find $\sim 0.29$ dex scatter around the star-forming main sequence for candidate blue nuggets below the threshold scale, which is consistent with temporary quenching as seen in simulations. A transitional population of green nuggets appears above the threshold scale. AGN also become more common in nuggets above this scale, and we see a likely AGN excess in nuggets versus comparably selected non-nuggets. Our results provide the first observational confirmation of the mass-dependent, AGN-mediated shift from cyclic quenching to halo quenching in nuggets.

*Unified Astronomy Thesaurus concepts:* Galaxy bulges (578); Galaxy evolution (594); Galaxy quenching (2040); Blue compact dwarf galaxies (165); Early-type galaxies (429); Elliptical galaxies (456); Galaxy formation (595); Lenticular galaxies (915)

*Supporting material:* machine-readable table

## 1. Introduction

The class of extremely compact quiescent galaxies now referred to as red nuggets was first identified at redshift $z > 1.6$ (Cimatti et al. 2004; Trujillo et al. 2006; van Dokkum et al. 2008). Red nuggets significantly modified the landscape of extragalactic astrophysics, as these galaxies are nearly $5\times$ smaller than typical galaxies of a similar stellar mass in today's universe and are rare at low $z$ (Naab et al. 2009; de la Rosa et al. 2016; Ferré-Mateu et al. 2017; Gao & Fan 2020).

Proposed red nugget progenitors were first found by Barro et al. (2013) and further confirmed by Williams et al. (2014) and Stefanon et al. (2013). These *blue nuggets* are nearly as compact as their red counterparts, but are starbursting instead of quenched. A toy model presented by Dekel & Burkert (2014) argued that the origin of these starbursting galaxies must be compaction events, i.e., externally fueled gas-rich processes that cause violent disk instability (VDI). VDI can result in angular momentum loss through gas collisions, which will lead to a direct increase of cold gas density in the galaxy's center and rapid star formation (SF). After the formation of a blue nugget, it can either quench temporarily or permanently, and permanent quenching will result in a red nugget. Blue nuggets are systems with highly centralized in situ growth, while permanently quenched red nuggets represent a stage just before ex situ growth via mergers. The toy model concludes that the ex situ growth of red nuggets is expected to create the massive elliptical population that can be seen in the local universe, but other studies have found that the in situ buildup of dense stellar cores in blue nuggets ultimately provides the seeds of all bulged galaxies, including lenticular and bulged spiral galaxies (de la Rosa et al. 2016; Penoyre et al. 2017; Costantin et al. 2020; Gao & Fan 2020).

Dekel & Burkert (2014) argue that the shutdown of cold accretion due to a hot halo—halo quenching—is likely one of the primary mechanisms that causes blue nuggets to permanently quench into red nuggets. Growing halos are expected to become hot over a critical halo mass range of $M_{\rm halo} \sim 10^{11.4}$ to $10^{12} M_\odot$, which drastically reduces cold gas accretion, and subsequently, quenches SF (Birnboim & Dekel 2003; Kereš et al. 2005; Dekel & Birnboim 2006). The key mass scale of $10^{11.4} M_\odot$ corresponds to the halo mass of a central galaxy with stellar mass $M_* \sim 10^{9.5}$ to $10^{9.7} M_\odot$, which Kannappan et al. (2013) call the *gas-richness threshold* scale due to the high frequency of gas-dominated galaxies below it. Halo mass $M_{halo} = 10^{12} M_\odot$ is referred to as the *bimodality* scale, as above that mass galaxies are typically composed of old stellar populations (e.g., Kauffmann et al. 2003; Baldry et al. 2004; Dekel & Birnboim 2006; Kannappan et al. 2013). High-$z$ nuggets ($z > 1.5$) can avoid quenching even above the bimodality scale through cold-in-hot accretion, as narrow dark matter filaments help cold streams pierce the shock fronts in hot halos (Dekel & Birnboim 2006).

Internal quenching mechanisms, such as stellar feedback and feedback from active galactic nuclei (AGN), have also been shown to help galaxies quench (Dekel & Silk 1986; Mac Low & Ferrara 1999; Springel et al. 2005; Kang et al. 2006; Martig et al. 2009; Hopkins et al. 2012). In simulations, Zolotov et al. (2015) found evidence that stellar feedback can help accelerate the blue-to-red nugget transition by expelling gas within the central bulge. In the case where a nugget is experiencing central bulge quenching alongside extended SF (also known as *inside-out quenching*), the combination results in what Dekel & Burkert (2014) refer to as green nuggets. Below the threshold scale, Zolotov et al. (2015) found that simulated nuggets may experience multiple inside-out quenching phases, but the lack of halo quenching prevents the nugget from permanently quenching (see also Tacchella et al. 2016b).

These predictions of halo mass quenching above the threshold scale and cyclic temporary quenching below the threshold scale have not yet been tested observationally. Most high-$z$ observational studies, such as those of Damjanov et al. (2009, 2011) and Barro et al. (2013), have focused on massive nuggets at or above the bimodality scale, so they cannot be used to confirm that halo quenching begins at the threshold scale. Other studies, such as those of Fang et al. (2013) and Wang et al. (2018), have been cited for having low-mass nuggets, but these samples focus on how surface mass density relates to quenching in the general galaxy population rather than in nuggets per se. Additionally, these samples do not extend below the threshold scale, where cyclic quenching is predicted. Palumbo et al. (2020) were the first to find that compact dwarf starburst (CDS) galaxies exist within the cyclic quenching regime at $z \sim 0$ and some of these CDS galaxies are likely low-$z$ blue nugget analogs, but they selected highly star-forming galaxies, thereby excluding quenching and quenched systems.

Some observational studies have probed the role of AGN feedback in nugget quenching. For example, Barro et al. (2013) found that massive compact star-forming galaxies host X-ray luminous AGN 30× more frequently than do noncompact massive star-forming galaxies. Whereas nugget simulations make clear predictions regarding the mass dependence of quenching, most simulations have not yet incorporated AGN feedback within their analysis and only speculate that AGN feedback likely plays a role in the blue-to-red nugget transition (Zolotov et al. 2015; Tacchella et al. 2016b). To assess nugget quenching mechanisms and their halo mass dependence, there is a need for a census of nuggets representing a wide range of stellar masses, star formation rates (SFRs), and AGN activity.

In this paper, we construct the first complete census of nuggets at all evolutionary stages within the volume- and mass-limited $z \sim 0$ REsolved Spectroscopy Of a Local VolumE (RESOLVE) survey (Kannappan & Wei 2008), the same survey used by Palumbo et al. (2020). Creating this census will allow us to answer the key question: *Do real nuggets show evidence of cyclic quenching below the threshold scale and permanent quenching above the threshold scale, as predicted by theory?* However, we first have to answer another question: *With respect to both past observational studies and the (theoretical) definition of nuggets as objects formed by compaction, how can we best select low-z nuggets?*

A challenge for our study is creating a data set of true nuggets (i.e., galaxies that formed via compaction). Observationally, high-$z$ nuggets are often selected on structural criteria that are not intended to be used for the low-mass regime where many low-$z$ nuggets are found (see de la Rosa et al. 2016 for a collection of selection criteria). Additionally, the frequency of nuggets is expected to decrease at low redshift while the frequency of compact non-nugget galaxies (e.g., dwarf ellipticals formed by galaxy harassment or compact ellipticals (cEs) formed by tidal stripping; see Moore et al. 1996; Norris et al. 2014; Ferré-Mateu et al. 2017) is expected to increase at low redshift. Finally, gas fractions are lower in the local universe, which may result in less dense nuggets (Dekel & Burkert 2014). For these reasons, simply replicating high-$z$ structural selection criteria is insufficient. Thus, *we must develop selection criteria, motivated by past observational studies and theory, to minimize nugget imposters among our nugget candidates.*

This paper is laid out as follows. In Section 2, we describe our data sets and the galaxy properties derived. In Section 3, we discuss the structural criterion for our initial selection of nuggets. In Section 4, we address our key question on the selection of nuggets, showing that additional environment-based selection criteria are required to isolate compaction-formed nuggets at low $z$. In Section 5, we review the general properties of our nugget candidates and confirm agreement with past observations and theoretical expectations. We go on to address our key question regarding quenching above and below the threshold scale. In Sections 6 and 7, we discuss the implications of our results and summarize our conclusions, respectively.

For our analysis, we adopt a standard $\Lambda$CDM cosmology with $\Omega_m = 0.3$, $\Omega_\Lambda = 0.7$, and $H_0 = 70$ km s$^{-1}$ Mpc$^{-1}$.

## 2. Data and Methods

### 2.1. RESOLVE Survey

To define our $z \sim 0$ nugget candidates, we start with the RESOLVE survey (Kannappan & Wei 2008). RESOLVE is a volume-limited census of stars, cold gas, and dark matter that covers >50,000 Mpc$^3$ over two equatorial strips. RESOLVE-B spans R.A. 22–3 hr and decl. $-1°.25$ to $+1°.25$, while RESOLVE-A is within R.A. 8.75–15.75 hr and decl. 0°–5°. RESOLVE galaxies are also required to have a group redshift of 4500–7000 km s$^{-1}$, thus creating two volumes defined by their R.A., decl., and redshift. RESOLVE-A is complete down to a selection limit at $M_r = -17.33$, which corresponds to the Sloan Digital Sky Survey (SDSS) apparent magnitude survey limit of 17.77 at the outer redshift boundary of RESOLVE. RESOLVE-B is complete down to a deeper selection limit at $M_r = -17.0$, due to repeat observations by SDSS (Eckert et al. 2015, hereafter E15). Throughout this study, we refer to the base sample of galaxies from which we identify nuggets as the parent survey. The parent survey is the complete, luminosity-limited RESOLVE survey, defined by the volumes just described along with an absolute $r$-band magnitude floor $M_r \leqslant -17.33$ (or $M_r \leqslant -17.00$ in RESOLVE-B). This complete RESOLVE parent survey contains 1453 galaxies.

## 2.2. Custom Photometry

We used custom-reprocessed multiwavelength photometry from E15 with minor updates from Hutchens et al. (2023). As detailed in E15, SDSS data were reprocessed using the improved background subtraction method of Blanton et al. (2011) and a combination of three methods of galaxy magnitude extrapolation. This resulted in brighter magnitudes and bluer colors when compared to the SDSS catalog. E15 used custom-reprocessed near-ultraviolet (NUV) from deep (>1000 s) GALEX observations (Morrissey et al. 2007) available for the entire RESOLVE footprint. Near-infrared (NIR) magnitudes were derived from the Two Micron All Sky Survey (2MASS; Skrutskie et al. 2006) and UKIDSS (Hambly et al. 2008). Using optical annuli created during the processing of $gri$, E15 successfully measured many more NIR magnitudes than are available within the 2MASS/UKIDSS catalogs. All the magnitudes we use are corrected for Milky Way extinction determined from the dust maps of Schlegel et al. (1998). Internal extinction and $k$-corrections are also applied as described below.

## 2.3. Stellar Masses, SFRs, Gas Masses, and AGN

Stellar masses were computed using a Bayesian spectral energy distribution (SED) fitting code, described in Kannappan et al. (2013) and last modified for E15. This SED fitting code also returns magnitudes that are $k$-corrected and corrected for internal extinction. SFRs are then derived from the UV data outputs from the SED fitting, custom mid-IR photometry derived from Wide-field Infrared Survey Explorer data (M. S. Polimera et al. 2024, in preparation), and the SFR prescription from Jarrett et al. (2012).

Gas masses derived from the 21 cm line come from Arecibo and the Green Bank Telescope, with some measurements taken from ALFALFA (Stark et al. 2016). Over 94% of RESOLVE has 21 cm detections or strong $3\sigma$ upper limits ($M_{HI} \leqslant 0.05$–$0.1 M_*$). The gas data were modified in Hutchens et al. (2023) to provide a best gas mass estimate by combining clean or successfully deconfused detections, strong upper limits, and photometric gas fraction estimates for weak upper limits or detections that could not be successfully deconfused.

AGN classifications for RESOLVE galaxies come from M. S. Polimera et al. (2024, in preparation), which extends the Polimera et al. (2022) classifications. Polimera et al. (2022) identified AGN using diagnostic plots (Baldwin–Phillips–Terlevich (BPT); Baldwin et al. 1981, VO; Veilleux & Osterbrock 1987) based on optical emission line fluxes (e.g., N II, O I, S II) with a particular focus on identifying AGN that hide in dwarf star-forming galaxies. This work revealed a new class of previously missed AGN (SF-AGN) that are mostly in dwarfs. SF-AGN do not register as AGN in the BPT plot, due to their low metallicity, but do show up in the VO plots. M. S. Polimera et al. (2024, in preparation) have added more AGN detected using mid-IR color and using updated BPT analysis. Thus, our AGN inventory is optimized for analyzing AGN in both dwarf and giant galaxies.

## 2.4. Environment Metrics

Environment metrics come from Hutchens et al. (2023), which uses a four-step group-finding algorithm to determine galaxy groups. This algorithm offers improved completeness and halo mass recovery when compared to the typical friends-of-friends group-finding method. Halo abundance matching was used to estimate the group halo mass for RESOLVE groups, including solitary galaxies (groups of one). The central is defined as the galaxy with the brightest absolute $r$-band magnitude within a group, while satellites are defined as the other galaxies within a group. We note that the halo masses used in this study are not subhalo masses, but rather group halo masses (including groups of one member).

## 2.5. PyProFit: Structural Parameters

The RESOLVE database contains structural parameters (e.g., effective radius, axial ratio) that come from the photometry of E15, but we have remeasured these parameters after correcting for atmospheric blurring. RESOLVE galaxies lie between $z = 0.015$ and $0.023$ (corresponding to physical scales of 0.32–0.49 kpc arcsec$^{-1}$), so some of the smallest galaxies in RESOLVE may be subject to atmospheric blurring comparable to their effective radius. To properly estimate the intrinsic sizes of RESOLVE galaxies, we used `PyProFit`, a Python wrapper to the light profile fitting algorithm `ProFit` (Robotham et al. 2017), on images from the Dark Energy Camera Legacy Survey (DECaLS; Dey et al. 2019) to create single-Sérsic profile models. Over 99% of galaxies in the RESOLVE survey have DECaLS images that can be used in `PyProFit`. The median full width at half-maximum (FWHM) for a point source in $r$-band DECaLS imaging is about 1″.2 (Dey et al. 2019), which is an improvement over the median SDSS $r$-band point source FWHM used in E15 of ∼1″.4.

We obtained background-subtracted $r$-band images and weight maps (∼3′ field of view and 0″.262 pixel scale) from DECaLS DR7. Some galaxies lie near the edges or corners of tiles, and in those instances, we obtained images and weight maps from the adjacent tile(s) and mosaicked the images using `SWarp` (Bertin 2010). We extracted segmentation maps to identify pixels in unique objects using the Python library of Source Extractor (Bertin & Arnouts 1996). Weight map images were used as uncertainty images.

We determined point-spread functions (PSFs) for each galaxy using stars within the field of view. We used `DAOStarFinder` from Stetson (1987) to detect isolated point sources and create $25 \times 25$ pixel cutouts that were resampled to $250 \times 250$ while conserving flux. We then performed a signal-to-noise ratio (S/N) weighted average on the PSFs and resampled back down to the $25 \times 25$ scale. We visually inspected all final PSFs and manually selected point sources to create PSFs for galaxies that failed the automated process for various reasons (e.g., artifacts, oddly shaped PSFs).

We performed single-Sérsic light profile fitting for all galaxies within the RESOLVE survey. For each galaxy, we provided five images as inputs: (1) an image of the galaxy, (2) an uncertainty image derived from the weight maps, (3) the segmentation map to indicate the regions to be fit, (4) the PSF image that `PyProFit` uses to blur the Sérsic model during the fitting process, and (5) the mask image to exclude specific regions from the fitting process. First, we centered the galaxy and cut the images to $4R_{90} \times 4R_{90}$, where $R_{90}$, the $r$-band 90% light radius, is from E15. Then, we created the mask image by using a Gaussian filter to blur the segmentation map. We next masked pixels belonging to objects that are not the galaxy of interest. Ultimately, the segmentation map used as an input to `PyProFit` simply marked the entire field of view (minus the masked objects) as regions to fit.

`PyProFit` accepts initial guesses for *x*- and *y*-position, magnitude, effective radius $R_e$, Sérsic *n*, position angle, axial ratio, and boxiness parameter of the Sérsic model. For the *x* and *y* positions, we provided initial estimates centered in the cutout, with bounds of $\pm 10$ pixels from the center of the galaxy. For the initial magnitude estimate, we used the existing apparent *r*-band magnitude derived by E15 with bounds of $\pm 1.5$. The $R_e$ from E15 was used as an initial guess for the `PyProFit` $R_e$ with a lower bound of $0.25 \times R_e$ and an upper bound of $2 \times R_e$, slightly favoring lower radii because `PyProFit` includes atmospheric blurring in the model fitting, whereas E15 did not. The Sérsic *n* initial estimate was 4 with lower and upper bounds of 1 and 12, respectively. All fits were allowed to range from $-180°$ to $180°$ in position angle and from 0.05–0.99 in axial ratio, with initial guesses for both parameters coming from E15. We allowed the boxiness parameter to range from $-1$ to 1 with an initial guess at 0. Robotham et al. (2017) Section 2.1 offers a more thorough explanation of the parameters and how they are used in creating light profiles. The fitting process itself is performed using the L-BFGS-B minimization algorithm within the `SciPy` package (Zhu et al. 1997). PyProFit returns a final value for each parameter, which can be used to create the full light profile. We show an example in Figure 1.

We visually inspected `PyProFit` models to flag failed or poor fits and to compare the models to other structural estimates. Depending on the specific failure mode, we refitted galaxies using different segmentation maps, initial parameter estimates, bounds, or minimization parameters to attempt to achieve a successful model. For RESOLVE galaxies that meet the luminosity-limited sample criteria (see Section 2.1), if the radius estimate from `PyProFit` and the radius estimate from E15 differed by over 50%, we conducted a visual check of the image with both $R_e$ values overlaid and flagged `PyProFit` models as *failed* where the `PyProFit` $R_e$ appeared incorrect relative to the image (note that sometimes the E15 $R_e$ was found to be incorrect instead when evaluated alongside the `PyProFit` $R_e$ and the image). Figure 2 illustrates the offset between successful `PyProFit` $R_e$ values and E15 $R_e$ values for galaxies within the parent RESOLVE survey. The values from E15 and `PyProFit` roughly agree, as the median value for $R_e > 10''$ hovers around unity. Below that, `PyProFit` returns lower $R_e$, due to modeling the seeing. The 123 galaxies in the parent survey without successful `PyProFit` models are included in our analysis using the seeing-uncorrected E15 $R_e$ (see histogram in Figure 2). While the median relation in Figure 2 could be used to correct the seeing effects for the 123 galaxies without Sérsic models, we chose to use the photometric $R_e$ from E15 for two reasons: (1) There is high scatter around the median relation, and (2) a significant fraction of failures correspond to either early-stage major mergers, whose radii are ill-defined, or faint irregular dwarfs, which are unlikely to overlap with nuggets.

We also compared `PyProFit` $R_e$ to $R_e$ from DECaLS DR9 Sérsic light profile modeling. Figure 3 shows the distribution of $R_e$ values for galaxies with successful models from both `PyProFit` and DECaLS. While the light profile models from DECaLS do not have the same by-eye visual quality control, we find that their structural estimates are roughly consistent with our `PyProFit` estimates.

Roughly 91% (1330/1453) of RESOLVE galaxies have successful $R_e$ estimates from `PyProFit`. The success rate of `PyProFit` for the RESOLVE parent survey is nearly identical to the ~91% success rate of `ProFit` (the R-variant of `PyProFit`) in Cook et al. (2019), in which the authors performed Sérsic light profile fitting on the xGASS survey. The flag for successful and failed `PyProFit` models and some `PyProFit` output parameters can be found in Table 1.

### 2.6. SF and Color Assignments

To track the evolutionary states of nuggets, we denoted each RESOLVE galaxy as being either a high-star formation (HSF), medium-star formation (MSF), or low-star formation (LSF) object. SF-based classification was performed using double-Gaussian fits to specific star formation rate (sSFR) versus stellar mass. As a consistency check, we also used $u - r$ to assign galaxies as belonging to the blue sequence, green valley, or red sequence, again based on double-Gaussian fits.

We used the Environmental COntext survey (ECO; Moffett et al. 2015, last updated in Hutchens et al. 2023) for the sole purpose of performing the double-Gaussian fits we used for color-sequence/SF category assignment. ECO was designed to have analogous data products (e.g., photometry, group finding) to RESOLVE while covering a roughly ~8× larger volume. In fact, RESOLVE-A is a subvolume within ECO. However, ECO lacks some of the high-quality data used in RESOLVE, such as uniform high-quality 21 cm and NUV data (see Sections 2.2 and 2.3). By virtue of its larger volume, ECO can provide more data points for the double-Gaussian fits. To define an approximately complete parent survey, we selected ECO galaxies with (1) a group redshift between 3000 and 7000 km s$^{-1}$, (2) reprocessed high-quality NUV data from GALEX (enabling accurate extinction corrections), and (3) stellar mass $M_* > 10^{8.9} M_\odot$ (following Eckert et al. 2016). This stellar mass-limited ECO survey allowed us to divide galaxies based on their color or SF in complete stellar mass bins down to $10^{8.9} M_\odot$. We created four stellar mass bins that range from $10^{8.9}$ to $10^{9.3} M_\odot$, $10^{9.3}$ to $10^{9.8} M_\odot$, $10^{9.8}$ to $10^{10.3} M_\odot$, and finally $10^{10.3} M_\odot$ and above.

To define SF categories, we fitted a double Gaussian over sSFR in each mass bin to separate the star-forming population and the quenched population. We then defined a *quenched* point and a *star-forming* point in each stellar mass bin. Both points are located at the same *x*-value, which is the median stellar mass within the bin. The *y*-value for the star-forming point is the sSFR where the star-forming Gaussian is $10\times \geqslant$ the quenched Gaussian, and the converse is true for the quenched Gaussian. We then separately fitted these star-forming points and quenched points across all stellar mass bins with tangent functions, as seen in Figure 4. Galaxies above the star-forming line are HSF, galaxies below the quenched line are LSF, and galaxies between the star-forming and quenched lines are MSF.

We note that the stellar mass-limited ECO survey allows us to create robust SF divisions that are free of bias. ECO and RESOLVE-A have the same luminosity completeness limit, and RESOLVE-B is slightly deeper (Section 2.1). We can create stellar mass-limited samples by requiring $M_* > 10^{8.9} M_\odot$ in RESOLVE-A and ECO, and $>10^{8.7} M_\odot$ in RESOLVE-B because at these stellar mass floors, nearly all the scatter in $M_*/L$ lies above the luminosity floors (Eckert et al. 2016). In the luminosity-limited RESOLVE survey, most of the galaxies that extend below the ECO/RESOLVE-A stellar mass limit are star-forming galaxies that qualify for the luminosity-limited survey due to their high luminosity for their mass, so fitting the divider in this regime would be subject to bias. Instead, we perform double-Gaussian fitting using the stellar mass-limited sample where the survey is complete for all $M_*/L$. This approach results in dividing lines with shallow slopes, which can be extrapolated down to lower masses with minimum bias. We use the larger ECO survey instead of just RESOLVE for the fitting solely for increased statistical power.

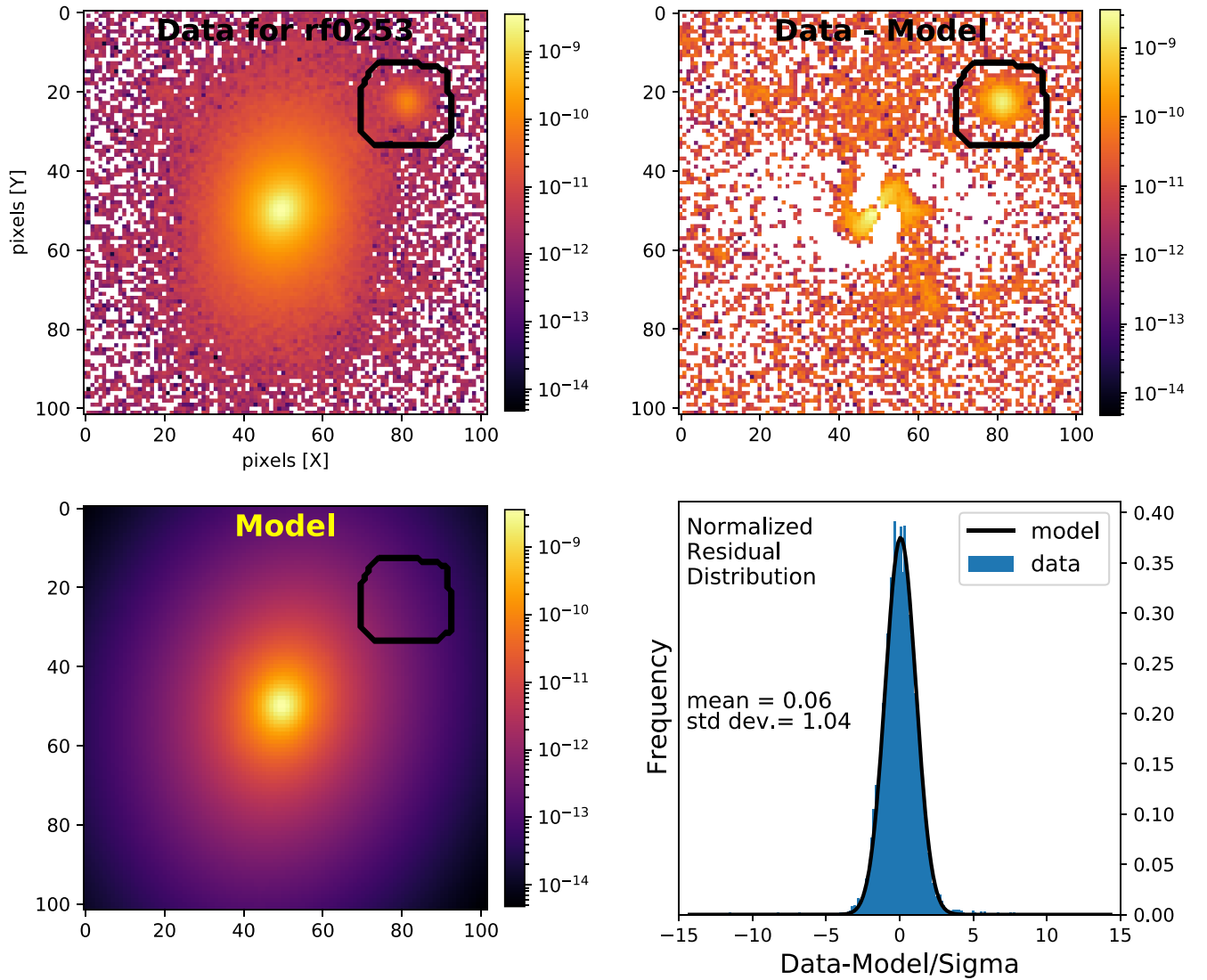

**Figure 1.** Data, model, and residual images for the `PyProFit` single-Sérsic model of RESOLVE galaxy rf0253. The *x*-axis and *y*-axis for the images are in units of pixels, and the color bar is in units of maggies. In this figure, the black contour denotes a region excluded from the fitting process. The bottom-right subplot is the distribution of the residuals of the fitted pixels normalized to the flux uncertainty value, sigma. A Gaussian model fit (black line and inset statistics) indicates residuals only slightly exceed the uncertainty. A spiral structure is revealed in the residual subplot. This figure replicates the analysis of `ProFit` models in Moffett et al. (2019).

For Figure 5, we similarly classified RESOLVE galaxies as blue sequence, green valley, and red sequence using a method that is nearly identical to the above method used for SF classification. We used the same four stellar mass bins and fitted a *blue* Gaussian and *red* Gaussian across extinction-corrected $u - r$ color. The green valley is assigned as the region between where the red Gaussian is $10\times \geqslant$ the blue Gaussian and vice versa. While tangent functions provide smoother divisions, we used line segments to create the $u - r$ divisions for ECO galaxies for consistency with Moffett et al. (2015) and Hoosain et al. (2024). As in these studies, our line segments extend horizontally both below the lowest-mass bin and above the highest-mass bin. Between the lines, galaxies are assigned as being within the green valley. Figure 5 shows $u - r$ versus stellar mass and where blue-sequence, green-valley, and red-sequence galaxies fall on the color sequences.

We tested both the SF division and color division by creating a star-forming main sequence (MS) offset plot in Figures 4 and 5. We defined the star-forming MS by iteratively fitting SFR against stellar mass while rejecting galaxies that fall >0.7 dex below the fit for each iteration, following Barro et al. (2017). Figure 4 shows the star-forming MS offsets for galaxies categorized by SF activity. We note that the separation of HSF galaxies from MSF/LSF galaxies in this study almost perfectly matches the star-forming MS offset of −0.7 used by Barro et al. (2017) to separate star-forming galaxies and quiescent

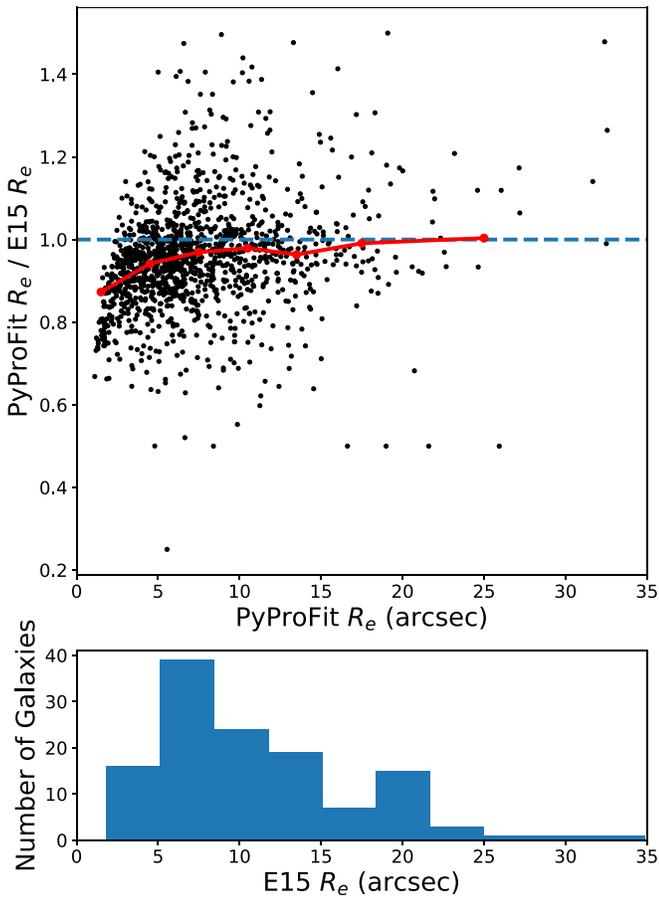

**Figure 2.** Top: `PyProFit` $R_e$ compared to E15 $R_e$ for galaxies with successful `PyProFit` models in the RESOLVE survey. A blue line at $y = 1$ marks unity for the ratio between the two $R_e$ measurements. The red line connects the median ratios in bins ranging from $3''-10''$ of `PyProFit` $R_e$. At the smallest radii, `PyProFit` estimates are typically smaller, as expected for seeing-deconvolved values. Bottom: distribution of E15 $R_e$ for the 123 galaxies that do not have successful `PyProFit` models.

galaxies. Analogously, Figure 5 shows star-forming MS offsets categorized by color sequence.

### 3. Initial Nugget Selection

We present our nugget selection criteria in two steps, beginning with an initial selection based purely on structure as done in most high-$z$ nugget studies, then adding secondary criteria necessary to eliminate nugget imposters in a later section (see Section 4.1).

In the following subsections, we discuss how nuggets have been identified in past studies and elucidate our approach to selecting nuggets based on both these past studies and differences to consider for local nuggets. For this paper, we will highlight two sets of nugget candidates: (1) SF divided, derived using the SF division in Section 2.6, and (2) color divided, derived using the color division in Section 2.6. We will show our findings primarily using the SF-divided nugget candidates, but we will also explicitly state whether there are any noteworthy differences when using the color-divided nugget candidates.

#### 3.1. Structure Criterion

A review of prior studies suggests nuggets have high central surface mass densities with typical effective radii of

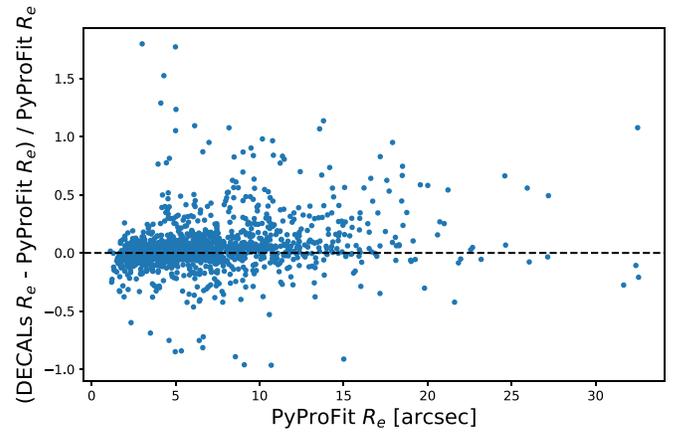

**Figure 3.** Comparison of `PyProFit` $R_e$ and DECaLS $R_e$ for galaxies in the parent survey that have both estimates. The dashed line marks where the fractional difference between estimates is zero. The median value for the data on the y-axis is 0.01 and the standard deviation is 0.24, suggesting strong agreement between the two radius estimates.

**Table 1**
RESOLVE `PyProFit` Parameters

| Parameter | Description |
|---|---|
| name | RESOLVE galaxy name |
| pfr50 | `PyProFit` effective radius |
| pfflag | `PyProFit` flag where $1/0/-1$ indicates an acceptable model/failed model/no model (e.g., due to missing inputs) |
| pfmag | `PyProFit` apparent magnitude |
| pfaxialratio | `PyProFit` projected axial ratio |

(This table is available in its entirety in machine-readable form.)

$\sim$1–2.5 kpc, sometimes reaching a maximum of $\sim$4 kpc (Damjanov et al. 2009; Barro et al. 2013; Dekel & Burkert 2014; Zolotov et al. 2015; Ferré-Mateu et al. 2017; Yıldırım et al. 2017; Martín-Navarro et al. 2019). Some nugget studies have not selected any structural metric, as red nuggets at high-$z$ can be effectively selected by identifying massive galaxies with old stellar populations (van Dokkum et al. 2008; Damjanov et al. 2009). Also, most nugget studies have used selection criteria specific to the high-mass regime, which prevents the selection of nuggets at all growth stages. The influential study of Barro et al. (2013) used log $(M_*/r^{1.5}[M_\odot \, \mathrm{kpc}^{-1.5}]) > 10.3$ to select red and blue nuggets above $M_* = 10^{10} M_\odot$ at high $z$, resulting in a sample of nuggets with a median effective radius of $\sim$1 kpc. A dwarf galaxy with $M_* \sim 10^{9.2} M_\odot$ would require a minuscule effective radius of $\sim$100 pc to be classified as a nugget under the structural selection criterion of Barro et al. (2013).

To create a structural selection criterion that varies with mass, we performed a forward fit to the stellar mass-effective radius relation (log $R_e$–log $M_\odot$) for LSF galaxies (or red-sequence galaxies for the color-divided parent survey). Candidate nuggets were selected to have a negative offset from the relation for LSF (or red sequence) galaxies (Figure 6). In other words, a nugget candidate at a given stellar mass must be smaller than the typical quenched galaxy at that same stellar mass, as dictated by the relation. The toy model of Dekel & Burkert (2014) suggests that nuggets will ultimately expand into quenched galaxies with compact bulges, so requiring our candidates to be denser than quenched galaxies will best capture nuggets before their ex situ accretion phase. Using the log $R_e$–log $M_\odot$ relation derived from LSF galaxies yields a maximum effective radius of ~4.2 kpc for our nugget candidates, in agreement with past studies. We also note that this criterion indirectly ensures that our nuggets host compact bulges when analyzed through a stellar surface mass density–stellar mass relation (as used in some studies): our relation using radius automatically constrains the nugget candidates to have higher effective stellar surface mass densities ($\Sigma_e$) than the typical red-sequence galaxy at a given stellar mass. Figure 6 illustrates our structural selection criterion, where the *initial* nuggets are the nugget candidates that pass this criterion.

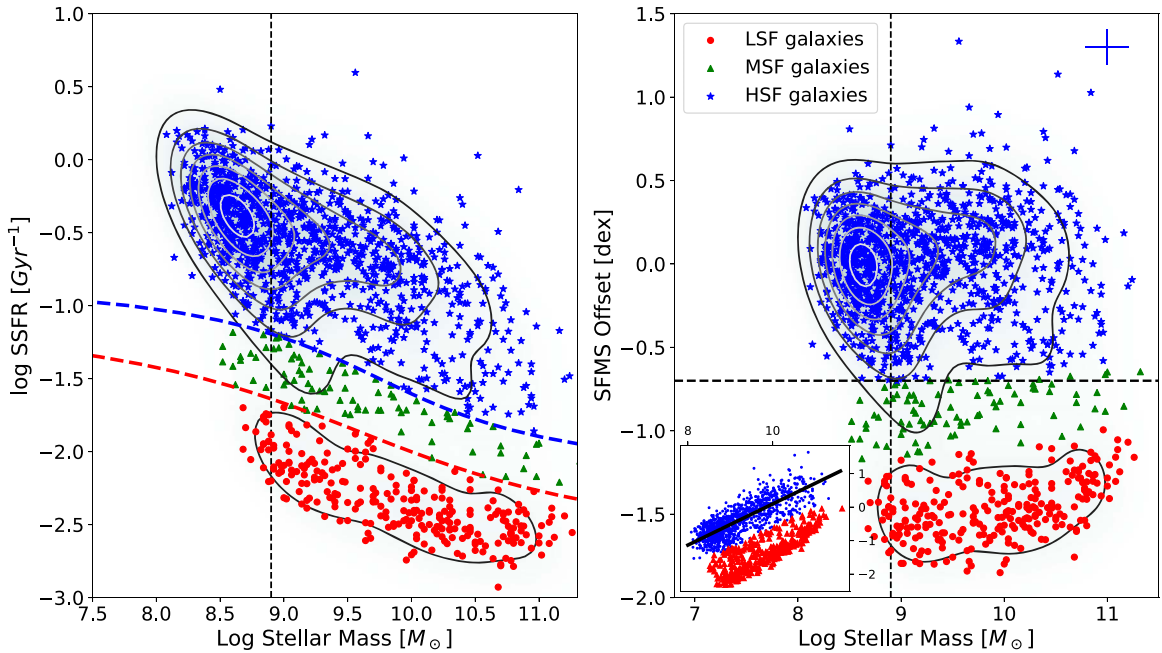

**Figure 4.** sSFR vs. stellar mass division of galaxies into SF categories (left) and correspondence to star-forming MS offsets (right). Point color and shape correspond to the assigned SF category. Vertical dashed lines in both panels represent the stellar mass limit used for ECO. The demarcation lines in the left panel were created using the stellar mass-limited ECO survey (Section 2.6). Points and contours were made using the smaller RESOLVE survey. The horizontal line in the right panel represents the −0.7 dex limit used to distinguish star-forming and quenched galaxies in Barro et al. (2017). Approximate errors can be seen in the top-right corner of the right panel. The inset shows log SFR vs. log stellar mass for the parent RESOLVE survey, with the star-forming MS (solid black line) calculated using the iterative fitting method described in Section 3.1 of Barro et al. (2017). The red triangles have star-forming MS offsets < −0.7 dex and, consequently, are rejected during the fitting process. Blue circles mark galaxies that were used to define the star-forming MS.

We have also selected an initial *ultracompact* nugget sample using a stricter structural criterion. Nuggets are known for their extremely high densities, and relic nuggets in the local universe that formed at high redshift are expected to be more dense than nuggets forming today (Trujillo et al. 2007; Dekel & Burkert 2014). While our initial standard nugget candidates were selected using the same strategy used in high-redshift studies (selecting all galaxies more compact than the quiescent galaxy mass–size relation, e.g., as in Barro et al. 2013), our initial ultracompact nugget candidates are defined by a line parallel to the quiescent galaxy mass–size relation that selects the most compact 25% of quiescent galaxies. This selection line can be seen as the red-dotted line in Figure 6. For this study, we will focus on the standard nugget sample and compare it to the ultracompact nugget sample when relevant.

### 3.2. Initial Nugget Candidates versus Similar Low-z Studies

The above selection criterion results in 291 SF-divided initial nugget candidates, where 147 are HSF nugget candidates, 21 are MSF nugget candidates, and 123 are LSF nugget candidates. For the color-divided parent survey, there are 328 color-divided initial nugget candidates, where 161 are red-sequence nugget candidates, 27 are green-valley nugget candidates, and 140 are blue-sequence nugget candidates. Our initial ultracompact nugget candidates are composed of 124 galaxies (61 LSF, 8 MSF, 55 HSF), which is less than half of the full set of initial nugget candidates and exactly half for initial LSF nuggets. We remind the reader that as SF (or color) is used to define the SF-divided (or color-divided) sample implicitly via the use of *only* LSF (or only red sequence) galaxies in the creation of the structure criterion, the SF-divided sample and color-divided sample do not contain the same nuggets. Below, we compare the density and size properties of the initial nugget candidates with past studies to ensure that they follow expectations.

Our initial nugget candidates display similar densities to compact central galaxies in Fang et al. (2013), which were used as examples of low-z nuggets in Dekel & Burkert (2014). The goal of Fang et al. (2013) was to better understand how inner density relates to quenching for central galaxies. Figure 6 of Fang et al. (2013) shows NUV-$r$ versus $\Sigma_{1\,\rm kpc}$ for central galaxies in different stellar mass bins starting at $M_* \sim 10^{9.75}\,M_\odot$. Their figure identifies crossover densities in each stellar mass bin, defined as the 20th percentile $\Sigma_{1\,\rm kpc}$ for quenched galaxies, above which galaxies are typically quenched. Since our LSF initial nugget candidates have median $M_* = 10^{9.89}\,M_\odot$, we use the two lowest-mass regimes (from $M_* = 10^{9.75}$ to $10^{10.25}\,M_\odot$) from Fang et al. (2013) to compare to our initial nugget candidates. These two mass bins have crossover densities that are roughly $\Sigma_{1\,\rm kpc} = 10^9\,M_\odot\,{\rm kpc}^{-2}$. $\Sigma_e$ is expected to be a factor of a few lower than $\Sigma_{1\,\rm kpc}$ for nuggets (Dekel & Burkert 2014). When selecting central LSF initial nugget candidates within the stellar mass range of $M_* = 10^{9.75}$ to $10^{10.25} M_\odot$ to directly compare to Fang et al. (2013), the crossover density is $\Sigma_e = 10^{8.59} M_\odot$ kpc$^{-2}$. The difference of 0.4 dex represents a factor of a few lower $\Sigma_e$ than $\Sigma_{1\,\mathrm{kpc}}$, showing agreement between the two studies.

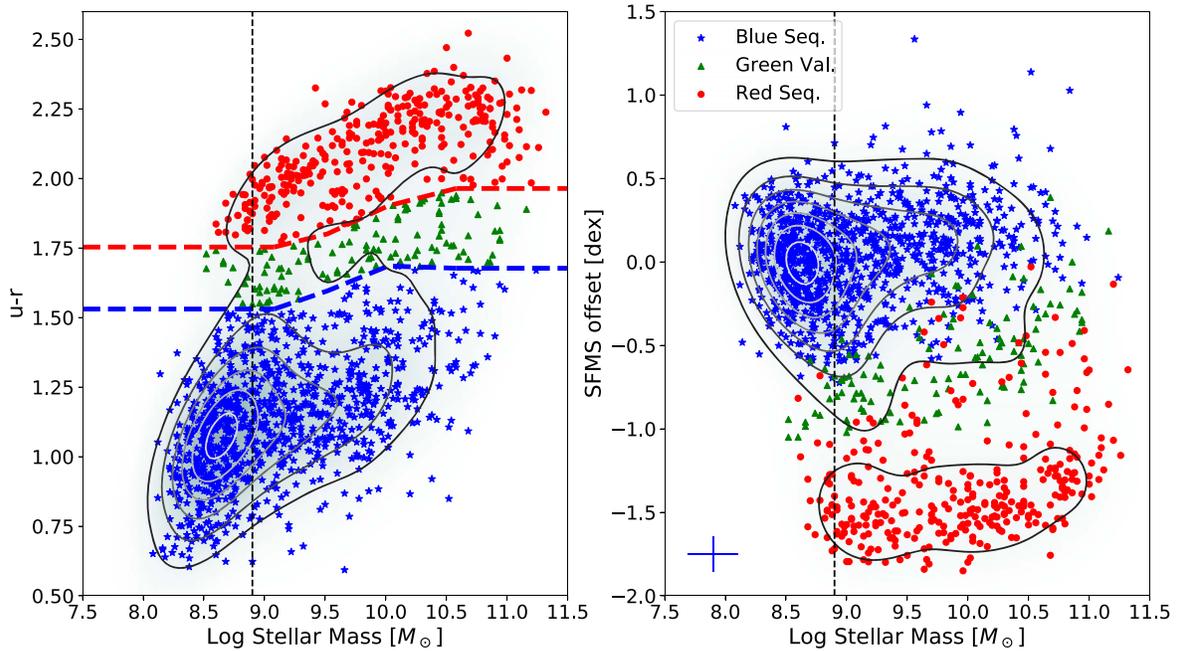

**Figure 5.** Division of galaxies into color categories using internal extinction-corrected $u - r$ vs. stellar mass (left) and correspondence to star-forming MS offsets (right). Points and contours in both subplots represent the RESOLVE parent survey. Point color and shape correspond to the assigned SF category. Vertical dashed lines in both panels represent the stellar mass limit used for ECO. Demarcation lines were created using the stellar mass-limited ECO survey (Section 2.6). Approximate errors can be seen in the bottom-left corner of the right plot.

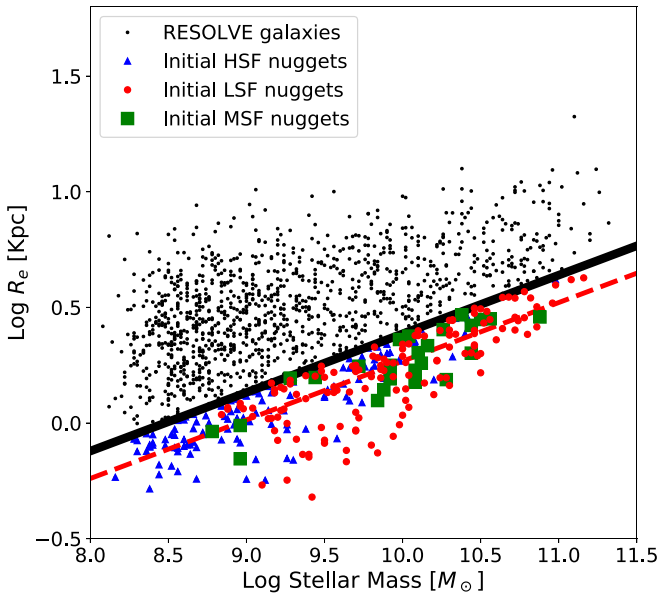

**Figure 6.** $R_e$ vs. $M_*$ for the parent RESOLVE survey. The black line represents the fitted LSF galaxy relation: the SF-divided initial nugget sample is defined to lie below the line (Section 3.1), while the rest of the RESOLVE survey (black dots) lies above the line. The red dotted line indicates the stricter selection criterion for the initial ultracompact nugget sample as described in Section 3.1.

We also compared the surface stellar mass density versus stellar mass relation of our initial SF-divided nugget candidates to the sample of another low-$z$ study that contains nugget candidates. Wang et al. (2018) evaluated the properties of a general sample of giant ($M_* > 10^{9.5} M_\odot$) galaxies and found starbursting compact systems with low gas content that appear to be nuggets in the act of quenching, as suggested by Palumbo et al. (2020), as well as quenched nuggets. The top-left subplot in Figure 4 of Wang et al. (2018) shows $\Sigma_{1\,\mathrm{kpc}}$ versus $M_*$ and their compact/extended division, which is defined as 0.2 dex below a linear fit in $\Sigma_{1\,\mathrm{kpc}}$–$M_*$ for quiescent galaxies. We replicate their compact/extended division by performing a linear fit in $\Sigma_e$–$M_*$ for LSF galaxies and then offsetting the fit by $-0.2$ dex. Based on this division, we find that all our initial nugget candidates are compact (see Figure 7). Our $\Sigma_e$–$M_*$ division is naturally $\sim$0.4 dex lower than the $\Sigma_{1\,\mathrm{kpc}}$ versus $M_*$ division used in Wang et al. (2018) at $M_* = 10^{9.5} M_\odot$, which is consistent with the expected factor of a few difference between $\Sigma_{1\,\mathrm{kpc}}$ and $\Sigma_e$ as per Dekel & Burkert (2014). An analogous result is found using the color-divided initial nugget candidates rather than the SF-divided initial nugget candidates.

The sizes of our initial nugget candidates are also within expectations. We find that the initial nugget candidates' median $R_e$ is 1.40 kpc. Using $M_* < 10^{9.5} M_\odot$ to identify dwarf nuggets, the median $R_e$ in this dwarf regime is 0.99 kpc, which is smaller than the median $R_e$ of 1.2 kpc for CDS galaxies in Palumbo et al. (2020). Above the dwarf regime, the median $R_e = 1.93$ kpc, lower than the 2.1 kpc median $R_e$ of the high-mass nuggets in Yıldırım et al. (2017). For the color-divided initial nugget sample, we obtain similar results: the median $R_e$, median dwarf $R_e$, and median giant $R_e$ are 1.47, 1.04, and 2.01 kpc, respectively. The median $R_e$, median dwarf $R_e$, and median giant $R_e$ for our initial ultracompact nugget candidates are 1.09, 0.80, and 1.49 kpc, respectively. Our ultracompact nuggets are significantly smaller than our standard nugget candidates by design, and our giant ultracompact nuggets have effective radii that are comparable to local relic nuggets in Ferré-Mateu et al. (2017).

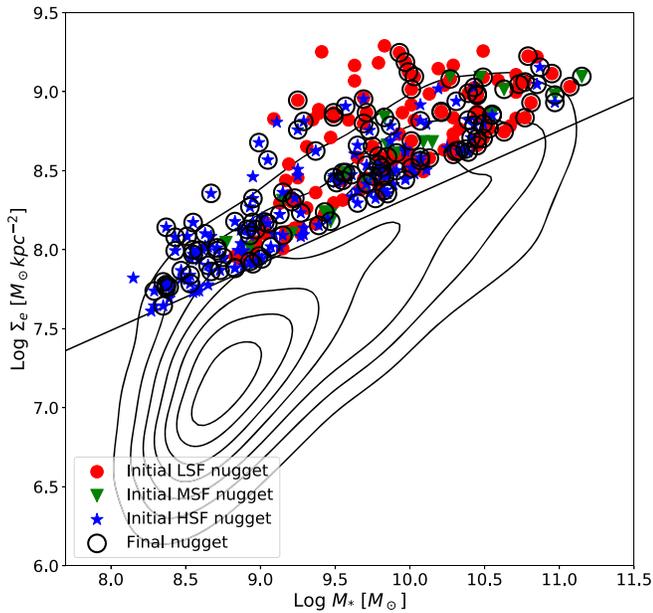

**Figure 7.** $\Sigma_e$ vs. $M_*$, illustrating agreement between our initial nugget sample and the compact galaxy selection of Wang et al. (2018). Red, green, and blue points represent LSF, MSF, and HSF nuggets, respectively, from our initial nugget sample defined in Section 3.2. Contours show the parent RESOLVE survey. Black circles represent the final nugget candidates defined in Section 4. The solid black line is adapted from the compact/extended divider used in Wang et al. (2018) to suit our use of $\Sigma_e$ rather than $\Sigma_{1\,\mathrm{kpc}}$.

## 4. Second-pass Improvements on the Nugget Candidates

In this section, we justify the need for a second set of selection criteria that focus on the environments of nugget candidates. Our environment-based selection criteria minimize the number of nugget imposters within our nugget sample. We then show that the final nugget sample agrees, within expectations, with past nugget studies.

A first look at the properties of the SF-divided initial nugget candidates shows all levels of SF activity at all stellar masses, including dwarfs ($M_* < 10^{9.5} M_\odot$) in the LSF region (Figure 8). Using the color-divided nuggets rather than the SF-divided nuggets still reveals a populated low-mass red sequence. Neither toy models nor simulations predict a dwarf, quenched nugget population (Dekel & Burkert 2014; Zolotov et al. 2015; Tacchella et al. 2016b). The primary mechanism that allows nuggets to permanently quench is expected to be hot halo quenching above a halo mass of $M_{\mathrm{halo}} \sim 10^{11.4} M_\odot$. This halo mass corresponds to a central galaxy stellar mass of $M_* \sim 10^{9.6} M_\odot$ (Eckert et al. 2016); therefore, lower stellar mass (dwarf) galaxies should not quench unless they are satellites in massive halos. Instead, dwarf centrals that experience compaction events should be mostly HSF nuggets. The simulations of Zolotov et al. (2015) and Tacchella et al. (2016b) do show temporary cyclic quenching below the threshold scale from internal quenching mechanisms, but this temporary quenching is not typically strong enough to move nuggets into the LSF regime. Observational studies of nuggets do not reach down to the dwarf regime for non-starbursting objects, so observational evidence of a dwarf LSF nugget population does not exist either. Thus, the apparent existence of dwarf, quenched "nuggets" suggests that these galaxies may not be true nuggets.

A possible source of nugget imposters among dwarf LSF nuggets is satellite galaxies, which may not have formed by compaction events but instead by various evolutionary mechanisms available to satellites. In Figure 9, we show the sSFR distributions of our SF-divided initial dwarf nugget candidates separated into centrals or satellites. A majority of dwarf centrals have high sSFRs, while dwarf satellites typically have low sSFRs. A Kolmogorov–Smirnov (KS) test suggests that these two populations are inconsistent with being drawn from the same population (p-value = 0.00065). This result implies that dwarf satellites and dwarf centrals have different sSFRs, due to environmental processes that can quench dwarf satellites. Below, we review various modes by which dwarf galaxies can become compact in appearance and classify them according to (a) whether they are compaction events, defined to involve gas-rich VDI as per Dekel & Burkert (2014) and (b) whether they typically affect centrals or satellites.

1. *Gas-rich mergers:* Gas-rich mergers can result in post-merger remnants with high central stellar densities (Mihos & Hernquist 1996; Hopkins et al. 2006; Zolotov et al. 2015), and are one of the primary compaction mechanisms in the toy model of Dekel & Burkert (2014). Gas-rich mergers should usually involve at least one central galaxy (Deason et al. 2014).

2. *Colliding gas streams:* As seen in both Tomassetti et al. (2016) and Zolotov et al. (2015), cold gas inflow can stimulate VDI and ultimately cause angular momentum loss and subsequent compaction. Filamentary flows travel toward the bottom of the potential well and are expected to fuel central galaxies.

3. *Harassment-induced starburst:* Repeat tidal interactions (harassment) in cluster environments can result in centralized starburst activity and compactness; however, this SF is not fueled by an external source of gas but rather by displacement of gas within the galaxy itself. Such harassment-induced starbursts are typically very short-lived ($<0.1$ Gyr) as they are not fueled by continuous gas inflow over a significant duration of time (Caldwell et al. 1999). In contrast, compaction events are expected to stimulate starbursts with lifetimes up to and beyond 1 Gyr, which is long enough to considerably increase the density of the bulge via in situ growth (Barro et al. 2013; Zolotov et al. 2015). Therefore, harassment does not involve enough gas or SF to qualify as compaction. By definition, harassment occurs only for satellites in dense environments.

4. *Tidal stripping:* Tidal stripping can remove all but the dense core of a galaxy. Stripping events are thought to be a common formation channel for compact galaxies in the local universe, such as cE galaxies (Huxor et al. 2011; Norris et al. 2014; Ferré-Mateu et al. 2018). Tidal stripping requires a more massive neighbor, so it applies primarily to satellites. Since it is not associated with cold gas inflow, galaxies formed by this mechanism are not nuggets, by definition.

5. *Pressure-confined starburst:* Du et al. (2019) designed a numerical simulation to explore the formation of cEs when satellites fall into massive host halos on elongated orbits. They found that while a large fraction of the satellite's gas is ram pressure stripped, the hot halo can exert pressure on the orbiting galaxy such that some of its cold gas gets compressed into the central bulge and fuels bursty SF. Since this mechanism only confines a galaxy's own gas into its center and does not involve continuous gas fueling, it does not involve enough gas to qualify as a compaction event. This mechanism only applies to satellites, as it requires a massive host halo to exert significant hot gas pressure along a highly elliptical orbit.

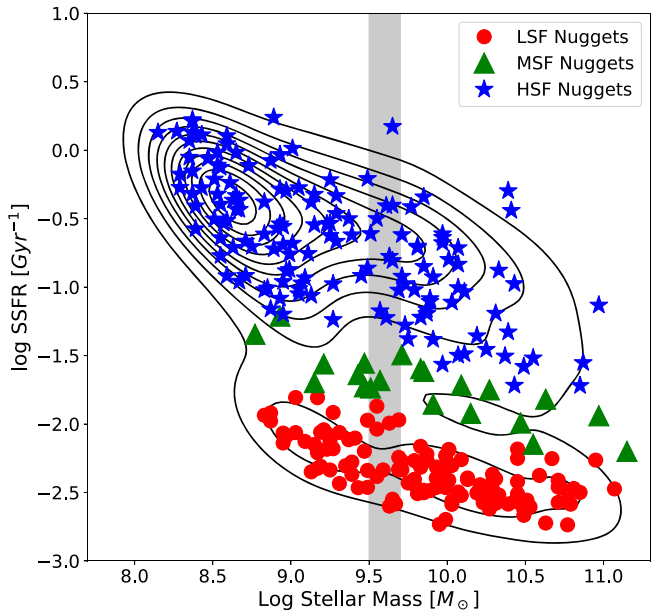

**Figure 8.** SF-divided initial nugget candidates in the sSFR vs. stellar mass plane, revealing a dwarf, quenched nugget population that is not predicted by theory, motivating our analysis of environmental selection criteria in Section 4. Marker color and shape are based on the assigned SF category from Section 2.6. Contours represent the parent RESOLVE survey. The shaded region corresponds to the typical stellar mass ($M_* \sim 10^{9.5}$ to $10^{9.7} M_\odot$) of a central galaxy at the halo mass threshold scale of $M_{\rm halo} = 10^{11.4} M_\odot$.

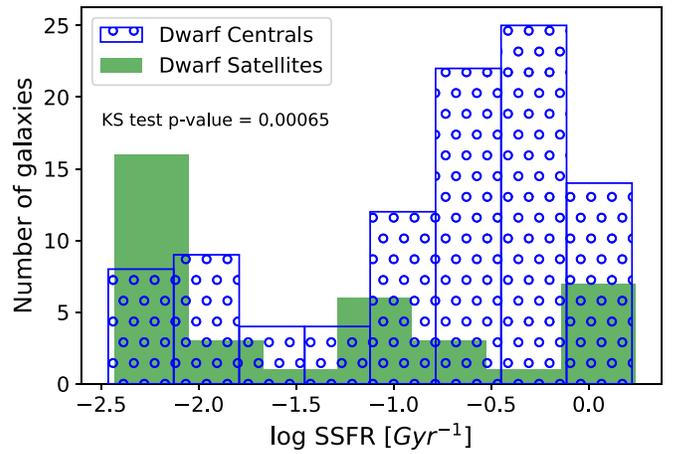

**Figure 9.** sSFR distributions of our SF-divided initial dwarf nugget candidates. Dotted and solid histograms show the color distribution of centrals and satellites, respectively. A two-sample KS test results in a *p*-value (inset) that suggests that these objects represent distinct populations.

While this is not an exhaustive review of all formation mechanisms for compact galaxies, it illustrates that true compaction events mostly involve centrals, while compact satellites are likely to be formed by non-compaction mechanisms. Below, we use this result to refine our selection criteria to better exclude nugget imposters.

*4.1. Additional Environmental Selection Criteria*

We now apply a second set of selection criteria, focused on environment, to minimize contamination by nugget imposters within our nugget sample. First, we require our nuggets to be centrals within their halo. Second, we require that our dwarf nuggets are not *flyby* galaxies, as such galaxies may have experienced recent harassment, stripping, or pressure-confined starbursts (Sinha & Holley-Bockelmann 2015). In a previous RESOLVE study, Stark et al. (2016) found that if centrals in halos below $M_{\rm halo} < 10^{12} M_\odot$ are gas-poor, then they commonly fall within $1.5\times$ the virial radius ($R_{\rm vir}$) of a nearby massive halo ($M_{\rm halo} > 10^{12} M_\odot$), implying that dwarf central galaxies within $1.5 \times R_{\rm vir}$ may still show signs of recent harassment or stripping despite not being formally part of the massive halo. Modifying the algorithm from Stark et al. (2016), we define flyby galaxies as galaxies in $M_{\rm halo} < 10^{12} M_\odot$ halos and within $1.5 \times R_{\rm vir}$ of a $M_{\rm halo} \geqslant 10^{12} M_\odot$ halo.

To recap, our final criteria for selecting nuggets at all evolutionary stages are:

1. The galaxy's effective radius must be offset toward smaller radii for its stellar mass than the $R_e$–$M_*$ relation for LSF (or red sequence) galaxies (see Section 3.1).

2. The galaxy must be a central.
3. Galaxies with $M_{\rm halo} < 10^{12} M_\odot$ must not be within $1.5\times$ the virial radius of a halo with $M_{\rm halo} \geqslant 10^{12} M_\odot$.

The final SF-divided nugget candidates, after incorporating both structural and environmental criteria, consist of 141 nuggets (89 HSF, 10 MSF, 42 LSF). The final color-divided nugget candidates consist of 157 nuggets (85 blue sequence, 14 green valley, 58 red sequence). There are 63 ultracompact final nugget candidates (25 LSF, 3 MSF, 35 HSF).

*4.2. Density, Size, and SF Activity of Final Nugget Candidates*

In Section 3.2, we showed that our initial nugget candidates display densities and sizes comparable to past studies. We can also confirm that the final nugget candidates also agree with previous studies. Figures 6 and 7 highlight the SF-divided final nugget candidates within the initial sample using black circles. Our final nugget candidates are structurally similar to the initial nugget candidates. Concerning the metric used to compare to Fang et al. (2013), the crossover density for our final LSF nugget candidates is $10^{8.61} M_\odot\,{\rm kpc}^{-2}$, which is similar to the crossover density for initial LSF nugget candidates of $10^{8.59} M_\odot\,{\rm kpc}^{-2}$. Regarding sizes, the SF-divided final nugget candidates have a median $R_e$ of 1.39 kpc, similar to the median $R_e$ of 1.40 kpc for the initial sample. The median $R_e$ values of the dwarf nugget candidates and the non-dwarf nugget candidates are 0.90 and 2.02 kpc, respectively. The color-divided final nugget candidates display similar densities and slightly larger sizes when compared to the SF-divided final nugget candidates. The median $R_e$, dwarf median $R_e$, and non-dwarf median $R_e$ for the color-divided final nugget candidates are 1.43, 0.94, and 2.16 kpc, respectively. Final ultracompact nugget candidates have sizes that are similar to those of initial ultracompact nugget candidates: the median $R_e$, dwarf median $R_e$, and giant median $R_e$ for our initial ultracompact nugget candidates are 1.13, 0.74, and 1.41 kpc, respectively.

We also find that our final nugget sample displays SF activity consistent with expectations. Figure 10 shows sSFR versus stellar mass and the star-forming MS offsets for our final nugget candidates. Compared to Figure 8, we find that the HSF, MSF, and LSF final nugget candidates can still be found at and above the threshold scale, but MSF and LSF nuggets are rare below the threshold scale. This follows expectations, as an LSF nugget population below the threshold scale is not expected in theory (e.g., Dekel & Burkert 2014; Zolotov et al. 2015). The color-divided final nugget sample also shows a significantly reduced number of red-sequence nugget candidates below the threshold scale.

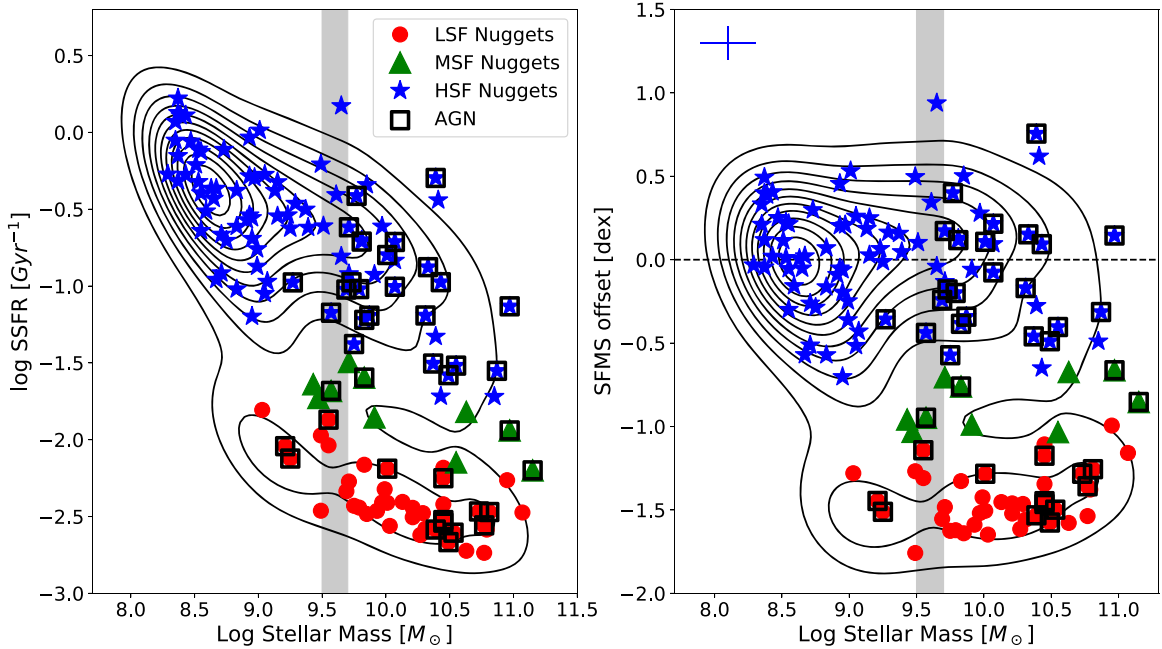

**Figure 10.** sSFR vs. stellar mass (left) and star-forming MS offsets vs. stellar mass (right) for our SF-divided final nugget candidates. In both panels, the contours are created using the parent RESOLVE survey. Both panels also show a shaded region, which corresponds to the typical stellar mass of a central galaxy with a halo mass at the threshold scale ($M_{\rm halo} \sim 10^{11.4}\, M_\odot$). Nuggets hosting AGN are denoted with an open square. Approximate errors in star-forming MS offsets and stellar mass can be seen in the top left of the right panel. The shortage of LSF and MSF nuggets below the stellar mass threshold scale suggests that permanent quenching occurs above the threshold scale for our final nugget candidates. The AGN frequency appears to increase above the threshold scale.

### 4.3. Compaction Event Origins

Our final dwarf HSF nugget candidates have plausible compaction event origins. Palumbo et al. (2020) identified 50 compact dwarf starburst (CDS) galaxies without the use of environment-based selection criteria within the RESOLVE survey and found photometric evidence of these galaxies having recent merger rates roughly 2× higher than a control sample. A spectroscopic analysis of a subset of CDS galaxies showed that 80% of them exhibited signatures of compaction via gas-rich mergers. These results are consistent with blue compact dwarfs being low-mass merger remnants (see also Bekki 2008; Stark et al. 2013). We found that 21 out of the 50 CDS galaxies in Palumbo et al. (2020) are within our SF-divided final nugget sample, which hints at our nuggets sharing the compaction event origins of CDS galaxies. The 28 remaining CDS galaxies from Palumbo et al. (2020) are either satellites/flybys or not compact enough to be in our final nugget sample. Similarly, 21 out of our 52 dwarf HSF nugget candidates are CDS galaxies in Palumbo et al. (2020). The other 31 dwarf HSF nugget candidates excluded from Palumbo et al. (2020) did not pass their starbursting criterion and/or morphology criterion (which selects on a parameter akin to Hubble type; see Kannappan et al. 2013). This difference is expected as dwarf blue nuggets are theorized to cyclically quench (e.g., Zolotov et al. 2015; Tacchella et al. 2016b), and we aim to select compact objects independent of their starbursting status or morphological classification. While a detailed analysis comparable to that of Palumbo et al. (2020) is beyond the scope of this study, we find it credible that our nugget candidates have formation channels similar to CDS galaxies.

### 4.4. Frequencies of Blue, Green, and Red Nuggets

We now examine the frequency of final nugget candidates to assess consistency with expectations for local nuggets. Theory predicts that nuggets have different evolutionary stories depending on whether they are below or above the threshold scale. Dwarf nuggets below the threshold scale are not expected to quench and may instead evolve into bulged disk galaxies, whereas permanent blue-to-red nugget evolution is theorized to occur above the threshold scale (see Section 1). Because of this, we focus on the frequency of HSF nugget candidates within the dwarf regime and the frequencies of LSF/MSF/HSF nugget candidates within the non-dwarf regime.

Our 53 dwarf HSF nugget candidates constitute ∼5.9% of the 902 dwarf galaxies within the luminosity-limited parent RESOLVE survey we used, suggesting that the frequency of our dwarf HSF nugget candidates is consistent with Palumbo et al. (2020) and Dekel & Burkert (2014). In particular, the model of Dekel & Burkert (2014) predicts that for a dwarf ($M_* = 10^{9.5}\, M_\odot$) galaxy in a halo at the threshold scale ($M_{\rm halo} = 10^{11.4}\, M_\odot$), the blue nugget frequency is ∼5% as discussed in Palumbo et al. (2020), who also found that the $z \sim 0$ CDS population constitutes roughly 5.3% of all dwarf galaxies in the RESOLVE survey. If we were to include all SF-divided dwarf nugget candidates, the frequency would become ∼6.6%. For the color-divided final nugget sample, blue-sequence dwarf nugget candidates and all dwarf nugget candidates constitute ∼6.3% and ∼7.2% of dwarf galaxies, respectively.

We also find that the volume number density of massive nuggets within our nugget census agrees with that of Hon et al. (2022). This study identified a sample of massive ($M_* > 10^{10.82} M_\odot$) nuggets within $z < 0.03$ and found that, depending on the nugget selection criteria used, the lower limit for the volume number density of these massive nuggets is $(0.12–1.72) \times 10^{-4}$ Mpc$^{-3}$. For our SF-divided nugget census, our volume number density for $M_* > 10^{10.82} M_\odot$ nuggets is $1.32 \times 10^{-4}$ Mpc$^{-3}$, consistent with Hon et al. (2022), although we note that there are only seven nuggets in our sample this massive. The color-divided final nugget candidates include nine nuggets this massive, yielding $1.70 \times 10^{-4}$ Mpc$^{-3}$.

The frequency of different nugget SF states above the threshold scale in the local universe varies somewhat with compactness. Final non-dwarf HSF and MSF nuggets make up 44/81 (54.3%$^{+5.4\%}_{-5.6\%}$) of our non-dwarf nugget sample, while LSF nuggets make up 37/81 (45.7%$^{+5.6\%}_{-5.4\%}$). For the final non-dwarf ultracompact nugget candidates, the frequency of HSF+MSF nuggets above the threshold scale is 15/36 (41.6%$^{+8.4\%}_{-7.8\%}$), while the LSF nugget frequency is 21/36 (58.3%$^{+7.9\%}_{-8.3\%}$). Thus, in contrast to standard nuggets, our ultracompact nugget candidates show a higher frequency of LSF nuggets compared to HSF+MSF nuggets, which may reflect a tendency of relic nuggets formed at high redshift to be more compact than more recently formed nuggets (see Barro et al. 2013).

Overall, these results appear consistent with a scenario in which low-$z$ nuggets above the threshold scale spend roughly half their life within compaction and quenching phases and the other half evolving out of the red nugget phase. Zolotov et al. (2015) found that simulated nuggets with $M_* \sim 10^{10} M_\odot$ (roughly the median stellar mass for our non-dwarf final nugget candidates) at $z \sim 2$ can take 1–3 Gyr to permanently quench after a compaction phase that lasts for $\sim$0.5–1 Gyr, totaling a lifetime of $\sim$1.5–4 Gyr. After the red nugget forms in the high-mass permanent quenching regime, Patel et al. (2013) suggest it will grow in effective radius by about 50% from z = 2 to z = 1 ($\sim$2.5 Gyr). Our nugget frequencies are plausibly consistent with these estimated lifetimes. However, Barro et al. (2013) found that the blue and red nugget frequencies between $z = 0.5$ and 3 depend heavily on redshift. We are not aware of a study that quantifies the frequency of blue and red nuggets at $z = 0$, so we can only compare to high-$z$ expectations.

## 5. Results

### 5.1. Halo Quenching as the Primary Driver of Nugget Quenching

We find that halo quenching is likely the dominant quenching mechanism responsible for the blue-to-red nugget transition above the threshold scale. Many studies have concluded that halo quenching is a requirement for permanent quenching (Kereš et al. 2005; Dekel & Birnboim 2006; Fang et al. 2013; Zolotov et al. 2015; Tacchella et al. 2016a; Martín-Navarro et al. 2019). Previous studies by Fang et al. (2013) and Martín-Navarro et al. (2019) found that while dense bulges are necessary for permanent quenching in local galaxies, they also hypothesize that shock heating in the halo must occur as this process is essential to suppress external gas accretion. These studies do not identify exact halo masses that correspond to quenching. In one of the few studies that explicitly states what halo masses correspond to permanent quenching in nuggets, Zolotov et al. (2015) reported that 70% of their simulated nuggets permanently quenched within the narrow halo mass range of $10^{11.4}$ to $10^{11.8} M_\odot$. In Figure 11, we show the distributions of gas-to-stellar mass ratio and star-forming MS offset with respect to the group halo masses for the SF-divided nuggets. The 84th percentile of halo masses for the HSF nuggets is $10^{11.90} M_\odot$, while the 16th percentile for LSF nuggets is $10^{11.45} M_\odot$. This result suggests that the halo mass range of $10^{11.45}$ to $10^{11.90} M_\odot$ is where HSF nuggets transition into LSF nuggets through permanent quenching, which is consistent with Zolotov et al. (2015). Figure 11 also shows that LSF nuggets exhibit low gas-to-stellar mass ratios compared to HSF nuggets, which is consistent with the theorized lack of cold-mode gas replenishment due to virial shocks. MSF nuggets also appear to be consistent with the permanent quenching scenario. While two MSF nuggets are seen within the errors of the threshold scale ($10^{11.35}$ and $10^{11.38} M_\odot$), most MSF nuggets are similar to LSF nuggets in that they exist primarily above the threshold scale and have low gas-to-stellar mass ratios. This similarity suggests that virial shocks in the halo are already suppressing external gas accretion in MSF nuggets. Analogous results for the color-divided sample can be seen, as the transition from blue-sequence nuggets to red-sequence nuggets occurs between halo masses of $10^{11.41}$ and $10^{11.75} M_\odot$. For the final ultracompact nuggets, the halo mass range is $10^{11.39}$ to $10^{11.64} M_\odot$, which is roughly consistent with the range for our final standard nugget sample and Zolotov et al. (2015).

We also find that the differences in the halo mass distributions of HSF, MSF, and LSF nuggets support the halo quenching theory for nuggets. Figure 12 shows the halo mass distributions of the nugget candidate subpopulations. The HSF nugget population decreases with increasing halo mass, while the LSF nugget population peaks just above the threshold scale, and then decreases. A two-sample KS test between these two populations returns a $p$-value of 0.00016, which confirms that LSF nuggets follow a different halo mass distribution than HSF nuggets. As for nuggets in the midst of quenching, MSF nuggets appear to be more similar to the LSF nugget population than the HSF nugget population. A two-sample KS test between MSF nuggets and HSF nuggets returns a $p$-value = 0.045, whereas a two-sample KS test between MSF nuggets and LSF nuggets yields a $p$-value = 0.37. These results agree with the notion that LSF and MSF nuggets exist primarily as a consequence of halo quenching, while HSF nuggets can exist both above and below the threshold scale. The color-divided nuggets show similar results: a two-sample KS test of the halo masses of the blue-sequence nuggets and the red-sequence nuggets yields a $p$-value = $7 \times 10^{-7}$.

A caveat of our selection process is that in requiring our nugget candidates to be central galaxies, we naturally enforce a tight stellar mass-halo mass relation (see Eckert et al. 2016). This selection makes it challenging to discern whether halo mass or stellar mass is a better predictor of permanent quenching in the final nugget candidates. However, we can analyze our initial nugget sample, which does not utilize environment-based selection, to probe whether stellar or halo mass more strongly predicts permanent quenching. Figure 13 shows the stellar mass–halo mass relation for our nuggets. For the final nuggets, the effects of group halo mass and stellar mass are intrinsically degenerate, but for the initial nugget candidates, the scatter only extends toward lower stellar mass at a given group halo mass. There are 33 LSF initial nugget candidates below the typical stellar mass of a central galaxy at the halo mass threshold scale ($M_* = 10^{9.5-9.7} M_\odot$; see Eckert et al. 2016), but after enforcing the environment-based selection criteria for the final sample, only three LSF nuggets remain below that same scale. This result suggests that halo mass drives permanent quenching more strongly than stellar mass within our nugget samples.

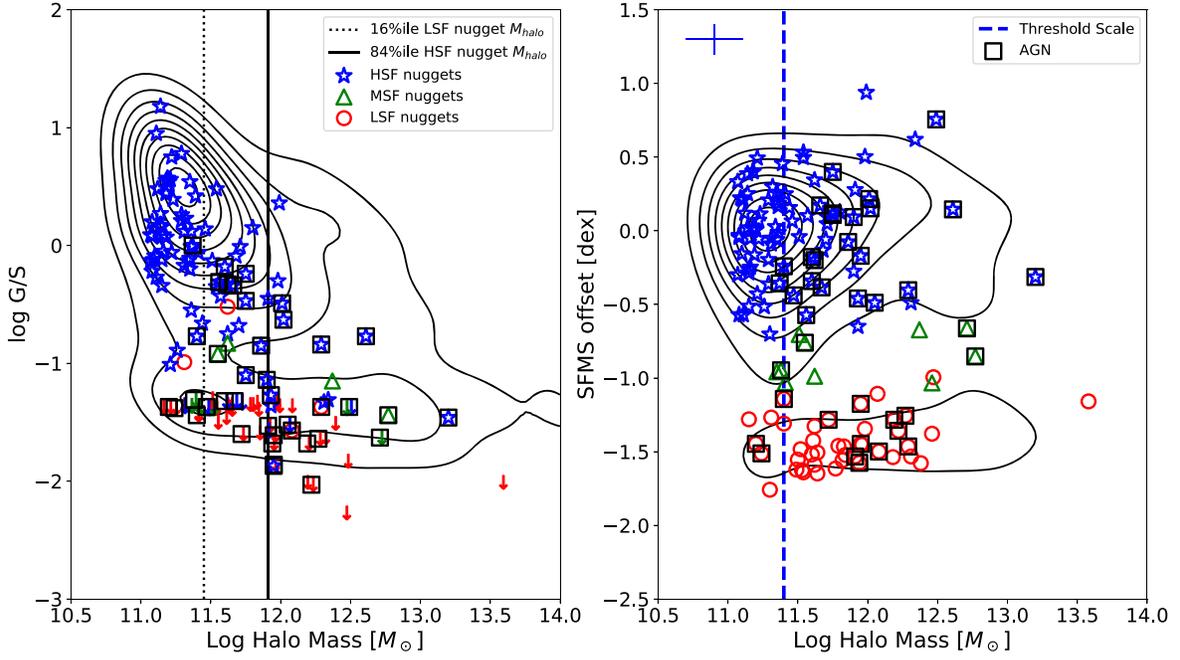

**Figure 11.** Gas-to-stellar mass ratios (left) and residuals from the star-forming MS (right), both as a function of halo mass, for our SF-divided final nugget candidates. Contours represent the parent RESOLVE survey. In the left panel, the dotted and solid vertical lines mark the 16th percentile halo mass for red nuggets and the 84th percentile halo mass for blue nuggets, respectively. These dashed lines show strong agreement with the halo mass regime where simulated nuggets in Zolotov et al. (2015) typically underwent permanent quenching. Arrows in the left panel represent the upper limits for gas measurements. In the right panel, the blue vertical line represents the halo mass threshold scale. HSF nugget candidates exist both below and above the threshold scale, while LSF and MSF nugget candidates primarily exist above the threshold scale. As in Figure 10, AGN are marked with open squares. Approximate errors can be seen in the top-left corner of the right panel.

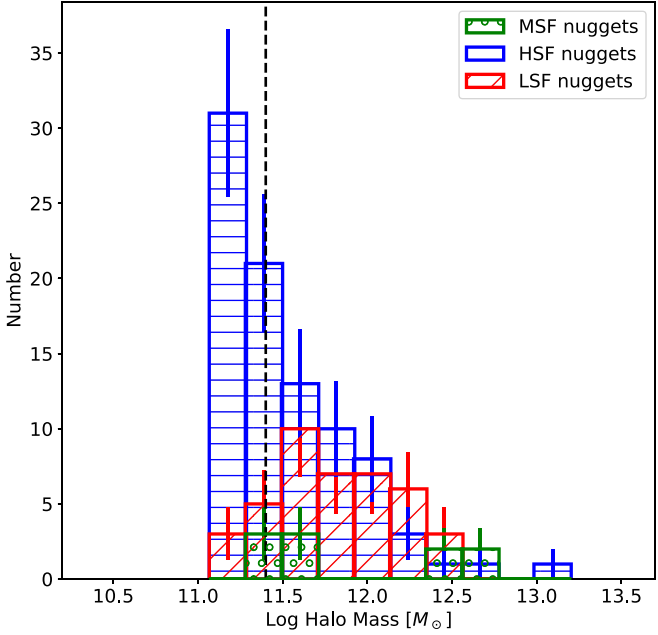

**Figure 12.** Histograms of the halo mass distribution of LSF, MSF and HSF nugget candidates. Error bars on the histograms represent $1\sigma$ Poisson errors. The black dashed line marks the threshold scale. KS tests between the three distributions suggest that MSF nuggets are more similar to LSF nuggets than to HSF nuggets. The sharp cutoff at the lowest mass is a consequence of halo abundance matching, which assumes a monotonic relationship between group halo mass and group luminosity. The lowest-mass halos are almost exclusively single-galaxy groups, so the luminosity floor of the parent survey effectively creates a halo mass floor at approximately $M_{\rm halo} = 10^{11} M_\odot$.

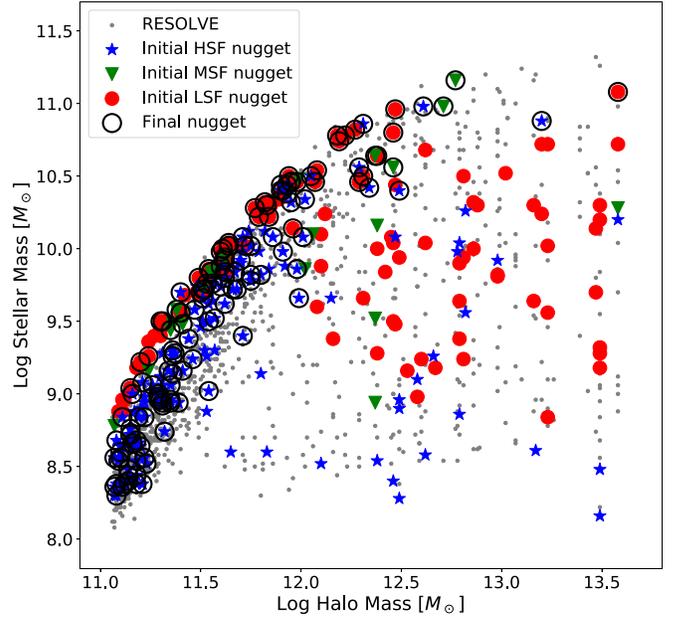

**Figure 13.** Stellar mass vs. group halo mass for the parent RESOLVE survey and nugget candidates. Red circles, green triangles, and blue stars represent initial LSF, MSF and HSF nugget candidates, respectively, as defined in Section 3.2. Black circles represent the final nugget candidates defined in Section 4. Gray points represent the rest of the parent RESOLVE survey. High-mass halos host low stellar mass nuggets that are quenched in the initial nugget sample, prior to our rejection of satellites, suggesting that halo mass, rather than stellar mass, drives quenching.

### 5.2. Temporary Quenching below the Threshold Scale

Our final nuggets are consistent with cyclic quenching below the threshold scale. Tacchella et al. (2016b) and Zolotov et al. (2015) found that cyclic quenching should result in a ∼0.27 dex scatter around the star-forming MS, with most nuggets not reaching the permanently quenched regime. Figure 10 shows the star-forming MS offsets for our SF-divided final nuggets. We find that the standard deviation of the star-forming MS offsets for the HSF nuggets below the threshold scale is 0.35 dex, or 0.29 dex when correcting for the typical star-forming MS measurement error of 0.2 dex. The corrected star-forming MS offset is consistent with the ∼0.27 dex scatter seen in simulations. We note that three of the seven MSF+LSF nuggets below the threshold scale host AGN, which may hint at them possibly experiencing very strong cyclic quenching due to AGN feedback. If we include them within the scatter calculation, the corrected standard deviation of star-forming MS offsets below the threshold scale is 0.42 dex. For the color-divided sample, the corrected standard deviation of star-forming MS offsets for blue-sequence nuggets is 0.32 dex, or 0.42 dex when including the AGN-hosting red-sequence/green-valley nuggets below the threshold scale. Our ultracompact nuggets show similar results: the corrected star-forming MS scatter for HSF nuggets below the threshold scale is 0.28 dex. Only one final LSF nugget (and one final red-sequence nugget) is a likely case of permanent quenching below the threshold scale, as the others either have halo masses within the errors of that scale or host an AGN.

### 5.3. AGN Frequency

Above the threshold scale, we find the AGN frequency is marginally higher in nuggets than in comparably selected non-nuggets, with the largest difference seen for HSF nuggets. Defining non-nuggets as galaxies in the parent RESOLVE survey that are not final nuggets and that are comparable to nuggets in being non-satellites/non-flybys (Section 4.1), we find that above the halo mass threshold scale, the nugget AGN frequency is 36/87 (41.3%$^{+5.3\%}_{-5.1\%}$, with $1\sigma$ binomial errors), whereas the non-nugget AGN frequency is 94/337 (27.9%$^{+2.5\%}_{-2.4\%}$). When restricting our analysis to HSF galaxies, we find that the AGN frequencies of nuggets and non-nuggets are 21/41 (51.2%$^{+7.7\%}_{-7.7\%}$) and 72/277 (26.0%$^{+2.7\%}_{-2.6\%}$), respectively. To ensure that our findings are not purely driven by stellar mass, we also evaluate the AGN frequency within the narrow stellar mass range of $M_* = 10^{9.5}$ to $10^{10} M_\odot$, just above the halo mass threshold scale. The AGN frequency within this stellar mass range is 13/36 (36.1%$^{+8.3\%}_{-7.5\%}$) for nuggets and 17/91 (18.7%$^{+4.4\%}_{-3.8\%}$) for non-nuggets. We find similar results for the color-divided nuggets. For the ultracompact nuggets, the AGN frequency above the threshold scale is 13/36 (36.1%$^{+8.3\%}_{-7.5\%}$), which is consistent within the errors with standard nuggets, albeit with large uncertainties due to small number statistics. While not statistically definitive, these results agree with past studies. Barro et al. (2013) found evidence of AGN being common in nuggets, stating that roughly 30% of their high-z blue nuggets hosted luminous X-ray AGN. Wang et al. (2018) used a sample of compact and extended star-forming galaxies at low z and found that compact star-forming galaxies are more likely to host Seyfert AGN than extended star-forming galaxies; the difference was strongest at stellar mass $M_* \sim 10^{10} M_\odot$.

Between the nugget subpopulations above the halo mass threshold scale, we find that the AGN frequency is consistent with AGN feedback playing a role in permanent nugget quenching. The AGN frequency of HSF, MSF, and LSF nuggets is 21/41 (51.2%$^{+7.7\%}_{-7.7\%}$), 3/8 (37.5%$^{+17.5\%}_{-14.8\%}$), and 10/35 (28.6%$^{+7.0\%}_{-8.2\%}$), respectively. Thus, as SF decreases, so does the likelihood of a nugget candidate hosting an AGN. Since both AGN and star-forming nuggets are fueled by cosmic gas, we expect AGN frequency to be highest within the HSF galaxy population (where gas fractions are high) and lowest within the LSF galaxy population (where gas fractions are low). Separately, elevated AGN frequencies coincide with galaxies above the threshold scale, suggesting that AGN likely play an important role in permanently quenching nuggets. As MSF nuggets become common above this mass scale, they may begin quenching due to both shock heating and AGN feedback, but the recent decrease of gas may have shut down some AGN, causing a drop in frequency compared to HSF nuggets. Figure 11 shows that MSF nuggets have depressed gas fractions, similar to LSF nuggets. Despite that fact, MSF nuggets have AGN frequency between HSF and LSF nuggets, implying that AGN in nuggets begin shutting down after gas fuel is expelled due to feedback. Thus, we conclude that the AGN frequency in HSF, MSF and LSF nuggets is consistent with AGN feedback expelling gas in permanently quenching nuggets above the threshold scale.

## 6. Discussion

As a reminder, the primary questions we began with were: (1) How can we best define low-z nuggets at all evolutionary stages? (2) Do real nuggets show evidence of cyclic quenching below the threshold scale and permanent quenching above the threshold scale, as predicted by theory? Our results have answered these questions, with some nuances, as detailed below.

### 6.1. Nuances of Identifying Nuggets

We have found that local nuggets at all evolutionary stages and masses can be selected using mass-dependent structure and environment criteria. Concerning structure, it is important to consider how compact a galaxy is with respect to both mass and the quenched galaxy population. Some studies, such as that of Barro et al. (2013), have circumvented the mass consideration by restricting their analysis to a narrow mass regime. However, the criterion used in Barro et al. (2013) would systematically miss low-mass nuggets. Similarly, mass-independent structural selection criteria, such as the radius selection used in Buitrago et al. (2018) to select massive nuggets, do not consider the mass dependence on nugget structure. If a flat radius selection criterion were applied to the RESOLVE survey, it would be too restrictive at higher masses and too inclusive at lower masses. Furthermore, compaction events are expected to be less intense at low redshift, creating blue nuggets that are not as compact as high-z blue nuggets (Dekel & Burkert 2014; Zolotov et al. 2015). Even in the case of selecting ultracompact nuggets, we find that our results are mostly consistent with those for standard nuggets. Ultracompact nuggets seem to contain a higher number of quenched objects than standard nuggets, which may be due to relic nuggets preferentially passing the stricter mass–size criterion. An elevated relic nugget fraction in our ultracompact nugget sample is expected as our ultracompact selection criterion is nearly identical to the criterion used to identify $z = 0$ relic nuggets in Flores-Freitas et al. (2022). As for how nuggets quench, ultracompact nuggets share a similar story to standard nuggets. The halo mass range for permanent quenching and the star-forming MS scatter linked to cyclic quenching in the ultracompact nugget sample are consistent with both standard nuggets and theorized expectations (Zolotov et al. 2015; Tacchella et al. 2016b). AGN fractions also agree within errors.

Environment-based selection criteria are necessary to exclude nugget imposters at low redshift. An example of a nugget imposter is a tidally stripped cE: such galaxies are found in the local universe with high central densities, low SFRs, and stellar masses below $M_* \sim 10^{10} M_\odot$. Ferré-Mateu et al. (2018) found that in a sample of 25 cEs in the local universe, 21 of them showed evidence of being the product of tidal stripping. Norris et al. (2014) also concluded that the majority of local cEs are formed in dense environments where tidal stripping is common. Therefore, identifying true nuggets is best achieved by selecting candidates in isolated environments where tidally stripped cEs are less common. Despite this, both Ferré-Mateu et al. (2018) and Norris et al. (2014) found subpopulations of putative cEs that appear to have formed via dissipative events, which would make them nuggets. While it is possible that we are excluding a unique subset of true nuggets with formation channels that depend on being satellites in dense environments, our nugget census is purer with the rejection of satellites and flybys, allowing us to reach accurate conclusions on how true nuggets quench and evolve.

Ultimately, identifying recent compaction is what truly defines nuggets, and it may be that both gas-rich mergers and colliding gas streams play significant roles in nugget formation in the local universe. In Section 4.3, we argue that our nuggets have compaction event origins based on similarity to and overlap with Palumbo et al. (2020). Palumbo et al. (2020) found that CDS galaxies (which have a strong correspondence with low-$z$ blue nugget candidates) are more likely to show recent merger evidence when compared to a control sample. Interestingly, four out of their seven CDS galaxies with spectroscopic data show some minor axis rotation, which can be evidence of formation via colliding gas streams (Tomassetti et al. 2016), but these candidates also display recent wet merger signatures.

In an attempt to better understand the likely formation channels for our low-$z$ nuggets, we tried to identify analogous nuggets in simulations so we could look into these galaxies' formation histories. We started with $z = 0$ snapshots of the Eagle simulation (Run ID = RefL0100N1504; Crain et al. 2015) and the IllustrisTNG simulations (Run ID = TNG-100-1; Pillepich et al. 2018). We identified a parent survey analogous to RESOLVE in each simulation by setting an absolute $r$-band magnitude floor $M_r \leqslant -17.33$. For the sake of comparison, we also defined a *modified* parent RESOLVE survey where both subvolumes are limited to this floor instead of the two separate floors described in Section 2.1. Figure 14 shows the sSFR distributions of the two simulated data sets and the modified parent RESOLVE survey.[8] Despite the apparent fundamental differences between these sSFR distributions, we performed nugget selection using the simulated data sets. As a proxy for identifying an LSF galaxy population, we calculated the ratio of LSF galaxies to all galaxies in the modified parent RESOLVE survey (310/1391, or 22%) and then assigned galaxies in the lowest 22% of sSFRs as LSF within the two simulated parent surveys. We next performed a forward fit to log $R_e$–log $M_\odot$ for the LSF galaxies. Finally, we identified our nugget candidates in the simulations by requiring that galaxies have a negative offset from the quenched mass–size relation and are centrals (flyby flags are not readily accessible within Eagle and IllustrisTNG). This returned 3834 IllustrisTNG nugget candidates and 2923 Eagle nugget candidates. We found that only 171 (4.5%) and 204 (7.0%) of $z = 0$ IllustrisTNG and Eagle nugget candidates had experienced mergers (with mass ratios $> 1/30$) within approximately the last gigayear. This result might imply that gas streams are a significant driver in local nugget formation.

---
[8] For TNG, effective radius, SFR, and stellar mass come from the `SubhaloHalfMassRadType`, `SubhaloSFR`, and `SubhaloMassType` parameters, respectively. For Eagle, effective radius, SFR, and stellar mass come from the `HalfMassRad_Star`, `StarFormationRate`, and `MassType_Star` parameters, respectively.

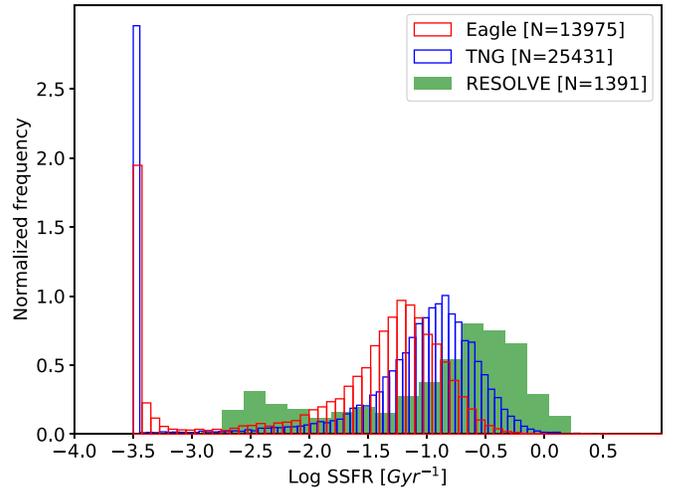

**Figure 14.** Distributions of sSFR for the RESOLVE survey (modified to use a floor of $M_r \leqslant -17.33$ in both subvolumes), IllustrisTNG simulation, and EAGLE simulation, where the simulations are also limited to $M_r \leqslant -17.33$. Log sSFRs were truncated to $-3.5$ Gyr$^{-1}$ to avoid unphysical sSFR values from simulations. SF activity appears inconsistent between the three data sets.

However, we also acknowledge that current cosmological hydrosimulations may be incompatible with a robust analysis of nugget formation. Our exploration into the Eagle and IllustrisTNG simulations showed significant differences from the observed galaxy population. For example, Figure 14 depicts distinct sSFR distributions for the three data sets. We find that 12% of HSF and MSF galaxies in the modified parent RESOLVE survey are more compact than the LSF mass–size relation, while the same percentages for IllustrisTNG and Eagle are much larger—22% and 37%, respectively. HSF and MSF galaxies in simulations often have densities similar to LSF galaxies, which conflicts with RESOLVE and other low-$z$ observational studies (e.g., Fang et al. 2013; Wang et al. 2018). Both structure and SF activity are vital for accurate nugget identification, so it may not be reasonable to use observational selection criteria to develop a census of nuggets with current cosmological hydrosimulation. The challenge of identifying nugget-like galaxies in cosmological hydrosimulations has been highlighted before. Flores-Freitas et al. (2022) explored IllustrisTNG to try to identify a population of relic nugget galaxies. Ultimately, they found no galaxies that could satisfy all the observational selection criteria used in Yıldırım et al.

(2017). Specifically, no simulated candidate nugget passed the mass–size criterion of Yıldırım et al. (2017). Thus, identifying nuggets in cosmological simulations to analyze their formation channels would have to take into consideration the significant differences between the general properties of simulated and observed galaxies.

### 6.2. The Crosstalk between Internal Quenching and Halo Quenching

Halo mass appears to be the strongest predictor of permanent quenching. We found that, regardless of the nugget classification method, most nuggets can only permanently quench once they reach the threshold scale. Below this scale, we saw evidence of cyclic quenching. This result agrees with the theoretical picture of virial shocks that turn on above the threshold scale and prevent abundant cold-mode accretion, consequently shutting down in situ stellar mass growth (Dekel & Birnboim 2006). Halo mass also seems to be a stronger predictor of permanent quenching than stellar mass is, given the number of quenched initial nuggets with low stellar mass in high-mass halos.

AGN feedback and stellar feedback may play a vital role in helping nuggets quench both above and below the threshold scale. High-$z$ studies have concluded that nugget quenching may be aided by both SF activity (e.g., Tacchella et al. 2016b; Zolotov et al. 2015) and AGN activity (e.g., Barro et al. 2013; Chang et al. 2017). Lapiner et al. (2023) used high-$z$ cosmological hydrosimulations to conclude that post-compaction AGN feedback can act as a source of quenching maintenance along with halo quenching. Given the apparent scarcity of AGN in the dwarf HSF nugget regime (Figure 11) despite our use of the most inclusive dwarf AGN inventory available (Section 2.3), stellar feedback may be primarily responsible for the presence of cyclic quenching. However, we do see two dwarf AGN among our few dwarf LSF nuggets, which is consistent with AGN feedback being important in stronger nugget quenching. Above the threshold scale, there may be an excess of AGN in nuggets when compared to nonnuggets. Other studies (e.g., Barro et al. 2013; Wang et al. 2018) have found that compact star-forming galaxies within large parent surveys (>3000 galaxies) exhibit significantly elevated AGN frequencies compared to extended star-forming galaxies. Our smaller parent survey shows consistent albeit weaker trends. We also see hints that the uptick in AGN frequency with respect to mass may occur at different masses for LSF versus HSF nuggets. Figures 10 and 11 show that star-forming nuggets have elevated AGN frequency above the threshold scale ($M_{halo} \sim 10^{11.4} M_\odot$, or for centrals: $M_* \sim 10^{9.6} M_\odot$), but quenched nuggets have increased AGN frequency closer to the bimodality scale ($M_{halo} \sim 10^{12} M_\odot$, corresponding to central galaxy mass $M_* \sim 10^{10.3} M_\odot$). Furthermore, the elevated AGN frequency in our nuggets between $M_* = 10^{9.5}$ to $10^{10} M_\odot$ is in agreement with $z = 0$ star-forming compact galaxies having a significantly higher Seyfert fraction than their extended counterparts at $M_* \sim 10^{10} M_\odot$ (Wang et al. 2018).

Ultimately, the crosstalk between halo quenching and internal quenching mechanisms (morphological quenching, AGN, and SF feedback) may be necessary to fully explain the blue-to-red transition in nuggets. Multiple studies have found a correlation between high surface mass density and quenching. That said, both internal and external quenching mechanisms were necessary to explain permanent quenching in Fang et al. (2013) and Martín-Navarro et al. (2019). Morphological quenching does not appear to play a significant role independent of other quenching mechanisms in our study, as only one LSF nugget appears to have a gas fraction similar to the typical HSF nuggets. However, evidence for morphological quenching may be degenerate with evidence for halo quenching, as halo quenching drives morphological changes (Kannappan et al. 2013; Moffett et al. 2015). Dekel & Birnboim (2006) also suggest that AGN feedback becomes more effective as halo mass increases. Thus, we cannot easily separate the relative importance of each quenching mechanism regarding the blue-to-red nugget transition.

For some nuggets, disk regrowth may also be able to occur during permanent quenching. Some studies have found that red nuggets above the threshold scale can be not only disky (e.g., van der Wel et al. 2011; Spiniello et al. 2021) but sometimes even diskier than blue nuggets (Barro et al. 2013). This implies that despite the shutdown of cold accretion, nuggets are still capable of building disks, perhaps during the green nugget phase. A study of nugget morphology at varying evolutionary stages would provide more details on exactly how nuggets engage in disk rebuilding.

### 7. Conclusion

We have created the first $z = 0$ census of nuggets at all evolutionary stages and robustly justified selection criteria that differ from those used at higher redshift. We have used this census to better understand the mass dependence and mechanisms of nugget quenching. The key questions we wanted to answer were:

1. With respect to both past observational studies and the (theoretical) definition of nuggets as objects formed by compaction, how can we best select low-$z$ nuggets?
2. Do our nuggets show signs of temporary cyclic quenching below the threshold scale and permanent quenching at and above the threshold scale, as predicted by theory?

In summary:

1. We carried out single-Sérsic light profile fitting using `PyProFit` and DECaLS DR7 $r$-band imaging (Robotham et al. 2017; Dey et al. 2019) and used the resulting light profiles to improve upon existing structural metrics for the $z \sim 0$ RESOLVE survey, a luminosity and volume-limited census of >1400 galaxies complete into the dwarf regime. We classified the evolutionary stages of our galaxies by assigning them as high-star formation, medium-star formation, or low-star formation objects. We equivalently classified galaxies as red-sequence, green-valley, or blue-sequence objects based on internal extinction-corrected $u - r$.
2. We selected nuggets at all stellar masses by requiring them to be offset to sizes smaller than the LSF (or red sequence) mass–size relation. This criterion alone creates a census of candidate nuggets with densities and sizes that agree with past studies, but we also employ an ultracompact nugget selection that identifies objects that are the most compact 25% of galaxies relative to the LSF mass–size relation.
3. Because $z \sim 0$ compact galaxies can form via environmental mechanisms such as harassment or tidal stripping,

our final nugget sample rejects satellite/flyby galaxies. These final standard nuggets are consistent with past theoretical and observational work on nuggets with respect to sizes (Section 4.2), densities (Section 4.2), and frequencies (Section 4.4).

4. The final nugget sample suggests that permanent quenching occurs within a halo mass range of $\sim 10^{11.45}$ to $10^{11.9} M_\odot$, consistent with the simulations of Zolotov et al. (2015). Below the threshold scale, blue nuggets have star-forming MS offset scatter $\sim 0.29$ dex, consistent with cyclically quenching nuggets in Zolotov et al. (2015) and Tacchella et al. (2016b), albeit we do see slightly larger offsets for a few low-mass quenched nuggets.

5. With the caveat that small number statistics may influence our conclusions on MSF nuggets, we find that HSF and MSF nuggets have statistically distinct halo mass distributions, while LSF and MSF nuggets show evidence of being drawn from the same halo mass distribution. All MSF nuggets are found either above the threshold scale or within the error of the threshold scale.

6. Above the threshold scale, we find some evidence that nuggets may host AGN at a higher frequency than comparably selected non-nuggets ($41.3\%^{+5.3\%}_{-5.1\%}$ versus $27.9\%^{+2.5\%}_{-2.4\%}$). The AGN frequencies in HSF, MSF and LSF nuggets are 21/41 ($51.2\%^{+7.7\%}_{-7.7\%}$), 3/8 ($37.5\%^{+17.5\%}_{-14.8\%}$), and 10/35 ($28.6\%^{+7.0\%}_{-8.2\%}$), respectively. Below the threshold scale, the role of AGN in low-mass nuggets is less clear in our sample, due to small number statistics.

7. Most of our results for ultracompact nuggets are similar to our standard nuggets, but our ultracompact nuggets contain a higher fraction of quenched objects ($58.3\%^{+7.9\%}_{-8.3\%}$ versus $45.7\%^{+5.6\%}_{-5.4\%}$), possibly reflecting an increased proportion of relic nuggets formed at high-$z$ among ultracompact nuggets.

8. We find that the Eagle and IllustrisTNG simulations show low rates of recent mergers for nugget-like galaxies (4.5% and 7.0%, respectively), with the caveat that we also confirm previous evidence (Flores-Freitas et al. 2022) that cosmological hydrosimulations are not yet producing realistic nuggets at $z = 0$.

This study is the first to provide a census of nuggets at all evolutionary stages within the local universe. The proximity of these nuggets is conducive to follow-up investigations aimed at further exploring their formation channels, structure, kinematics, and post-nugget evolution.

## Acknowledgments

We thank Adrienne Erickcek, Carl Rodriguez, Fabian Heitsch, Gerald Cecil, Abigail Dunnigan, and the referee for their valuable feedback and help regarding this project. This research has been supported by the National Science Foundation under award AST-2007351. D.S.C. acknowledges additional support from the Graduate Research Fellowship of North Carolina Space Grant and the Summer Research Fellowship by UNC-CH Graduate School.

*Software:* Matplotlib, a Python library for publication-quality graphics (Hunter 2007); SciPy (Virtanen et al. 2020); the IPython package (Pérez & Granger 2007); Astropy, a community-developed core Python package for Astronomy (Astropy Collaboration et al. 2013, 2018, 2022); NumPy (Harris et al. 2020); pandas (McKinney 2010, 2011).

## ORCID iDs

Sheila J. Kannappan https://orcid.org/0000-0002-3378-6551
Manodeep Sinha https://orcid.org/0000-0002-4845-1228
Michael L. Palumbo, III https://orcid.org/0000-0002-4677-8796
Kathleen D. Eckert https://orcid.org/0000-0002-1407-4700
Mugdha S. Polimera https://orcid.org/0000-0001-6162-3963